%% file: DeepLearningVis-VIS2017.tex
\documentclass[10pt,journal,cspaper, compsoc]{IEEEtran}
\ifCLASSOPTIONcompsoc
  \usepackage[nocompress]{cite}
\else
  \usepackage{cite}
\fi

\ifCLASSINFOpdf
  \usepackage[pdftex]{graphicx}
\else
  \usepackage[dvips]{graphicx}
  \DeclareGraphicsExtensions{.eps}
\fi

\usepackage[pdftex]{graphicx}
\usepackage{times}
\usepackage{paralist}
\usepackage{amsmath}
\usepackage{url}
\usepackage{cite}
\usepackage{stfloats}
\usepackage{array}
\usepackage{ragged2e}

\usepackage{bm}
\usepackage{amssymb}
\usepackage{verbatim}
\usepackage{multirow}
\usepackage{makecell}
\usepackage{ulem} 
\usepackage{stfloats}
\usepackage{color,microtype,hyphenat,balance}
\usepackage{graphicx,subfigure}
\usepackage{booktabs}
\usepackage{amsthm}
\usepackage{algorithmicx}
\usepackage{overpic}
\usepackage{rotating}
\usepackage{tikz}

\usepackage{xcolor}
\usepackage[english]{babel}
\usepackage{wrapfig}
\usepackage{ulem}
\usepackage[singlelinecheck=false]{caption}

\usepackage[ruled,vlined]{algorithm2e}
\DeclareMathAlphabet{\mathcal}{OMS}{cmsy}{b}{n}
\DeclareMathAlphabet{\mathcal}{OMS}{cmsy}{m}{n}

\newcommand{\mc}[1]{\textcolor{black}{#1}}

\newcommand{\doc}[1]{\textcolor{black}{#1}}
\newcommand{\kelei}[1]{\textcolor{black}{#1}}
\newcommand{\keleiRe}[1]{\textcolor{black}{#1}}
\newcommand{\keleiReRe}[1]{\textcolor{black}{#1}}
\newcommand{\newdoc}[1]{\textcolor{black}{#1}}
\newcommand{\todo}[1]{\textcolor{black}{#1}}
\newcommand{\jw}[1]{\textcolor{black}{#1}}
\newcommand{\mcrev}[1]{\textcolor{black}{#1}}
\newcommand{\jing}[1]{\textcolor{black}{#1}}
\newcommand{\jwReRe}[1]{\textcolor{black}{#1}}

\newcommand{\shixia}[1]{\textcolor{black}{#1}}
\newcommand{\xia}[1]{\textcolor{black}{#1}}

\newcommand{\B}{\mathrm{Beta}}

\newcommand{\pictag}[2]{$\textsf{{}#1}_\textsf{{}#2}$}

\newcommand{\vecz}[1]{\boldsymbol{z}_{{}#1}}
\newcommand{\caselayer}[1]{$\textsf{L}_\textsf{{}#1}$}
\newcommand{\casefm}[1]{$\textsf{F}_\textsf{{}#1}$}
\newcommand{\caseexpert}[1]{$\mathrm{E_{{}#1}}$}
\newcommand*{\circled}[1]{\lower.7ex\hbox{\tikz\draw (0pt, 0pt)%
    circle (.5em) node {\makebox[1em][c]{\small #1}};}}

\usepackage[bookmarks,backref=true,linkcolor=black]{hyperref} 

\ifCLASSOPTIONcompsoc
\else
\fi

\ifCLASSINFOpdf
\else
\fi

\begin{document}
%
\title{Analyzing the Noise Robustness of \\Deep Neural Networks}

\author{
Kelei~Cao, Mengchen~Liu, Hang Su, Jing Wu, Jun Zhu, Shixia~Liu
\IEEEcompsocitemizethanks{
\IEEEcompsocthanksitem K. Cao and S. Liu are with School of Software, BNRist, Tsinghua University. S. Liu is the corresponding author.
\IEEEcompsocthanksitem H. Su and J. Zhu are with Dept. of Comp. Sci. \& Tech.,  Institute for AI, THBI Lab, Tsinghua University.
\IEEEcompsocthanksitem M. Liu is with Microsoft; J. Wu is with Cardiff University.
}
\thanks{}}

%


\teaser{
\setcounter{figure}{0}
\centering
\vspace{-14pt}
  \scriptsize
  \begin{overpic}[width = 0.92\linewidth]{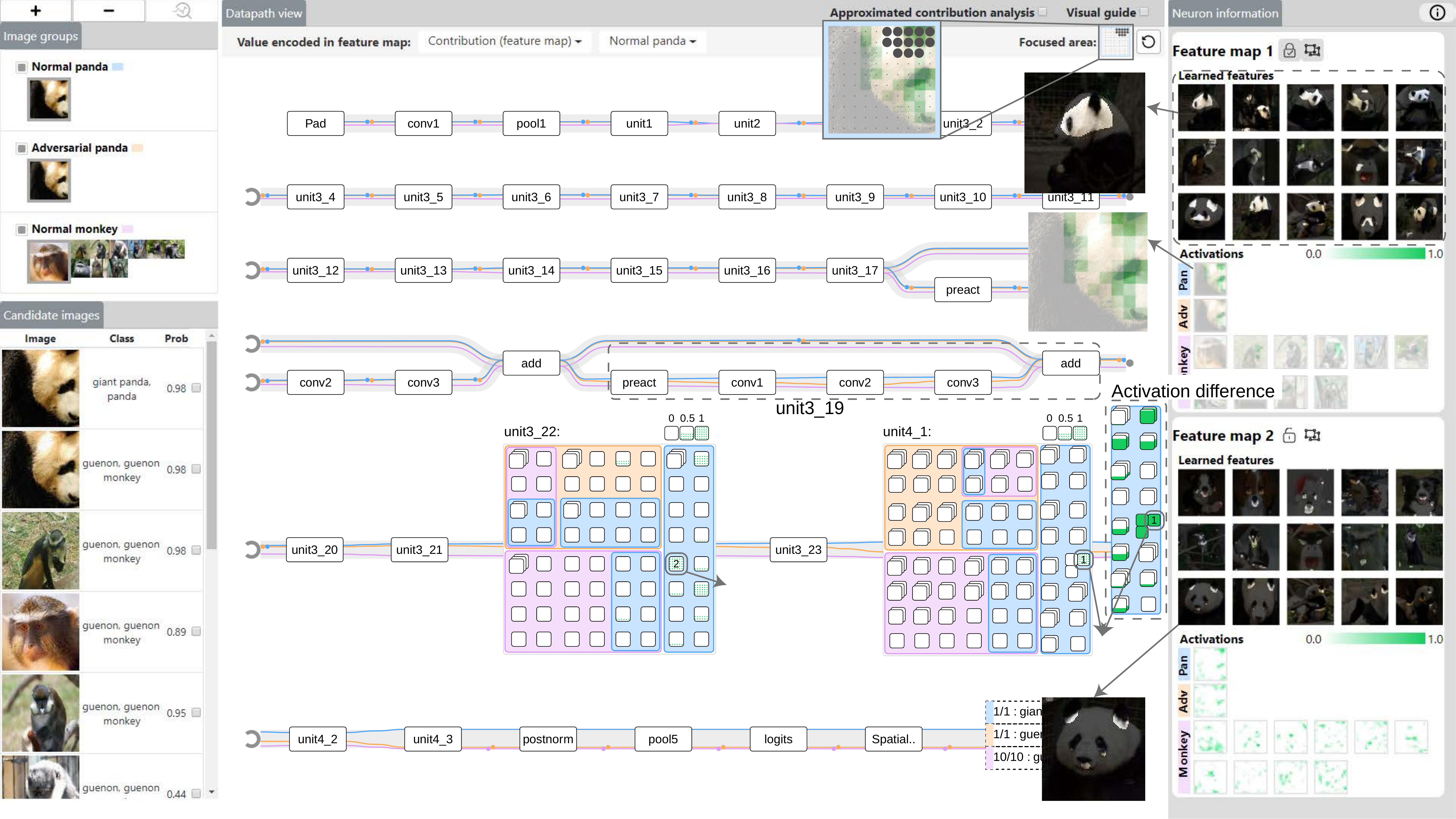}
    \put(8.7,51.2){$\textsf{I}_\textsf{1}$}
    \put(10.03,45.55){$\textsf{I}_\textsf{2}$}
    \put(9.4, 40.15){$\textsf{I}_\textsf{3}$}
    \put(32.5,11.9){$\textsf{L}_\textsf{A}$}
    \put(54,19.8){$\textsf{L}_\textsf{B}$}
    \put(58.5,11.7){$\textsf{L}_\textsf{C}$}
    \put(21,6.8){$\textsf{L}_\textsf{D}$}
    \put(29,6.8){$\textsf{L}_\textsf{E}$}
    \put(50,14.5){$\textsf{F}_\textsf{A}$}
    \put(75.2,11){$\textsf{F}_\textsf{C1}$}
    \put(96, 52){$\textsf{A}$}
    \put(69.1,44.2){$\textsf{B}$}
    \put(69.2,34){$\textsf{C}$}
    \put(64.8, 52.5){$\textsf{D}$}
    \put(70.2, 1){$\textsf{E}$}
    \put(2, -1.5){\textsf{(a) Input images}}
    \put(27, -1.5){\textsf{(b) Datapath visualization at the network- and layer-levels}}
    \put(82.5, -1.5){\textsf{(c) Neuron visualization}}
  \end{overpic}
     \vspace{2mm}
\setcaptionwidth{0.92\linewidth}
  \caption{
Explanation \newdoc{
}why an adversarial panda image is not classified as a panda.
The root cause is \jw{identified as} 
\newdoc{
}\jw{the neurons in the feature map $\textsf{F}_\textsf{A}$ 
\newdoc{
failing} to detect} 
the outline of the panda's ear ($\textsf{E}$) in the adversarial example,  
which further leads to the failure of detecting the panda's ear ($\textsf{B}$) in $\textsf{F}_\textsf{C1}$.\looseness=-1
}
\vspace{-6mm}
\label{fig:teaser-overview}
}

\IEEEcompsoctitleabstractindextext{%
\justify
\begin{abstract}
Adversarial examples, generated by adding small but intentionally imperceptible perturbations to normal examples, can mislead deep neural networks (DNNs)
\newdoc{
to make}
incorrect predictions.
Although much work has been done on both adversarial attack and defense,
\keleiReRe{a fine-grained understanding of adversarial examples is still lacking}.
To address this issue, we present a visual analysis method to explain why adversarial examples are misclassified.
The key is to compare and analyze the datapaths of \newdoc{both} the adversarial and normal examples. A datapath is a group of critical neurons
\newdoc{
along}
with their connections.
We formulate the datapath extraction as a subset selection problem and solve it by constructing and training a neural network.
A multi-level visualization consisting of a network-level visualization of data flows, 
a layer-level visualization of feature maps, 
and a neuron-level visualization of learned features\newdoc{, has been} designed to help investigate how datapaths of adversarial and normal examples diverge and merge in the prediction process.
\newdoc{A quantitative} evaluation and a case study 
\newdoc{were} conducted to demonstrate the promise of our method 
\newdoc{to explain} the misclassification of adversarial examples.\looseness=-1
\end{abstract}
\begin{IEEEkeywords}
Robustness, deep neural networks, adversarial examples, explainable machine learning.
\end{IEEEkeywords}}

\maketitle
{
\fontsize{10}{10} 
\input{introduction}
\input{related}

\input{system}

\input{datapath}

\input{visualization}

\input{application}

\input{discussion}

\input{conclusion}
}

\IEEEdisplaynotcompsoctitleabstractindextext

%
\IEEEpeerreviewmaketitle


%

 \ifCLASSOPTIONcompsoc
 \section*{Acknowledgments}
\else
   regular IEEE prefers the singular form
 \section*{Acknowledgment}
\fi

K. Cao and S. Liu are supported by the National Key R\&D Program of China (No. 2018YFB1004300) and the National Natural Science Foundation of China (No.s 61936002, 61761136020, 61672308).
H. Su and J. Zhu are supported by the National Key  R\&D Program of China (No. 2017YFA0700904), NSFC Projects (Nos. 61620106010, 61621136008), Beijing NSF Project (No. L172037), Beijing Academy  of Artificial Intelligence (BAAI), and the JP Morgan Faculty Research Program.

\ifCLASSOPTIONcaptionsoff
  \newpage
\fi



%
\bibliographystyle{IEEEtran}

\bibliography{reference}
\vspace{-6mm}
\begin{IEEEbiography}
[{\includegraphics[width=1in,height=1.25in,clip,keepaspectratio]{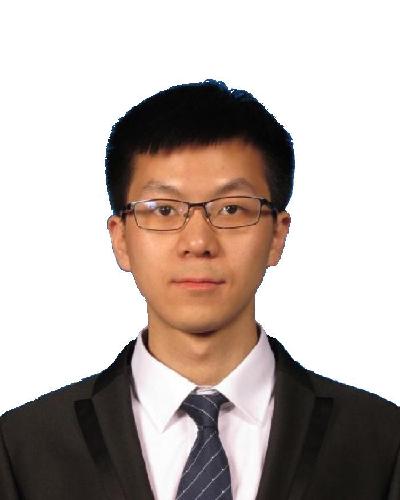}}]{Kelei Cao} is now a master student at Tsinghua University. His research interest is visual analytics for explainable deep learning. He received a BS degree in the Department of Computer Science and Technology, Tsinghua University.
\end{IEEEbiography}

\begin{IEEEbiography}
[{\includegraphics[width=1in,height=1.25in,clip,keepaspectratio]{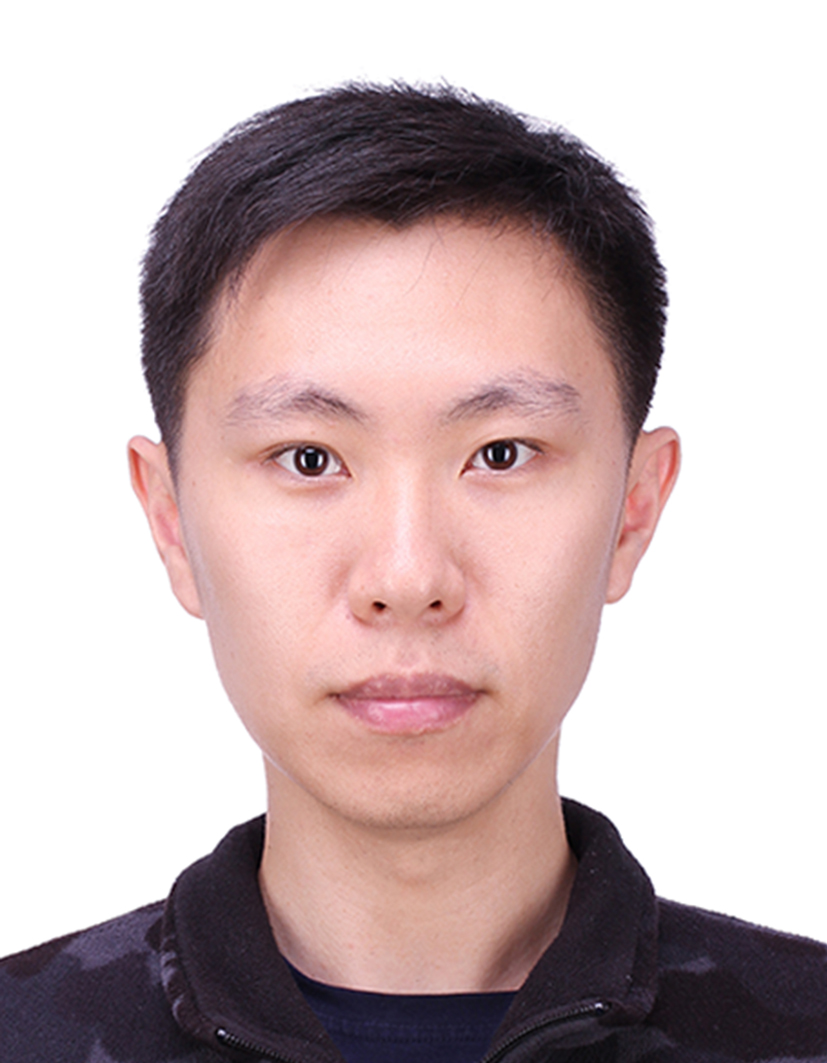}}]{Mengchen Liu} is a Senior Researcher at Microsoft. His research interest includes explainable AI and computer vision. He received a BS in Electronics Engineering and a Ph.D in Computer Science from Tsinghua University.
He has served as PC members and reviewers in various conferences and journals.
\end{IEEEbiography}

\begin{IEEEbiography}
[{\includegraphics[width=1in,height=1.25in,clip,keepaspectratio]{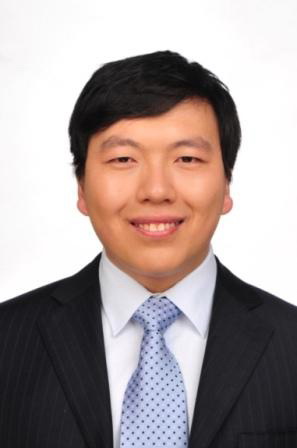}}]{Hang Su} is an assistant professor at Tsinghua University. He received his B.S, M.S., and Ph.D. Degrees from Shanghai Jiaotong University. His research interests lie in the development of computer vision and machine learning algorithms for solving scientific and engineering problems arising from artificial learning, reasoning, and decision-making. His current work involves the foundations of interpretable machine learning and the applications of image/video analysis. He has served as senior PC members in the dominant international conferences.
\end{IEEEbiography}

\begin{IEEEbiography}
[{\includegraphics[width=1in,height=1.25in,clip,keepaspectratio]{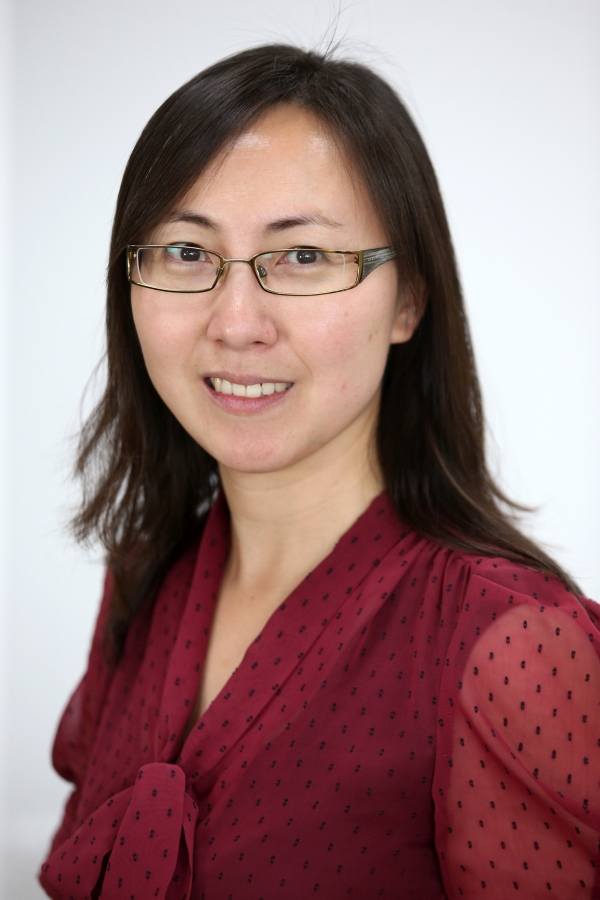}}]{Jing Wu} is a lecturer in computer science and informatics at Cardiff University, UK. Her research interests are in computer vision and graphics including image-based 3D reconstruction, face recognition, machine learning and visual analytics. She received BSc and MSc from Nanjing University, and PhD from the University of York, UK. She serves as a PC member in CGVC, BMVC, etc., and is an active reviewer for journals including Pattern Recognition, Computer Graphics Forum, etc.
\end{IEEEbiography}

\begin{IEEEbiography}
[{\includegraphics[width=1in,height=1.25in,clip,keepaspectratio]{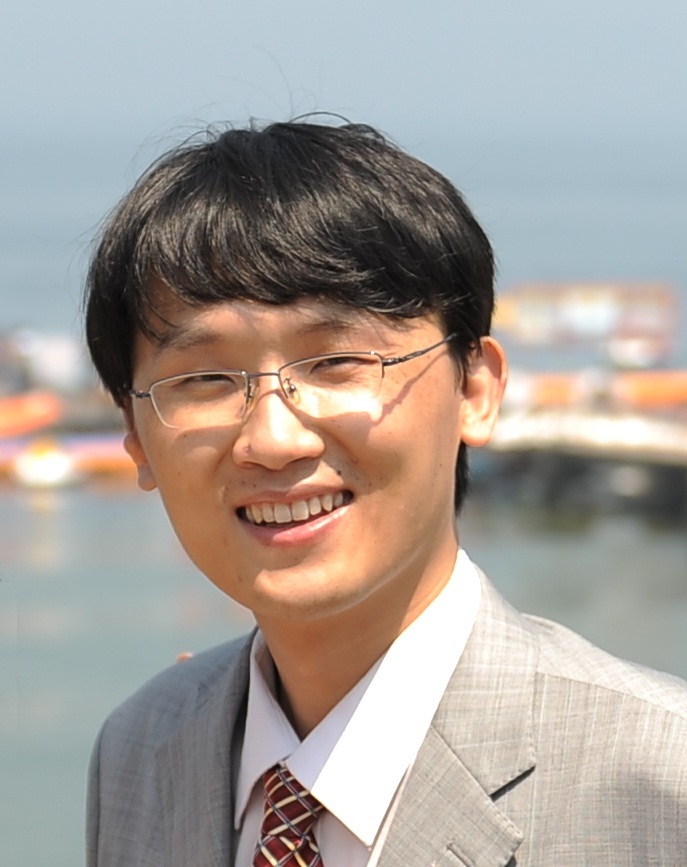}}]{Jun Zhu} received his BS and PhD
degrees from the Department of Computer
Science and Technology in Tsinghua
University, where he is currently a professor. He was an adjunct faculty and postdoctoral fellow in the Machine Learning Department, Carnegie Mellon University.
His research interests are primarily
on developing statistical machine learning
methods to understand scientific and engineering
data arising from various fields. He regularly serves as Area Chairs at prestigious conferences, such as ICML, NeurIPS. He is an associate editor-in-chief of IEEE TPAMI.
\end{IEEEbiography}

\begin{IEEEbiography}
[{\includegraphics[width=1in,height=1.25in,clip,keepaspectratio]{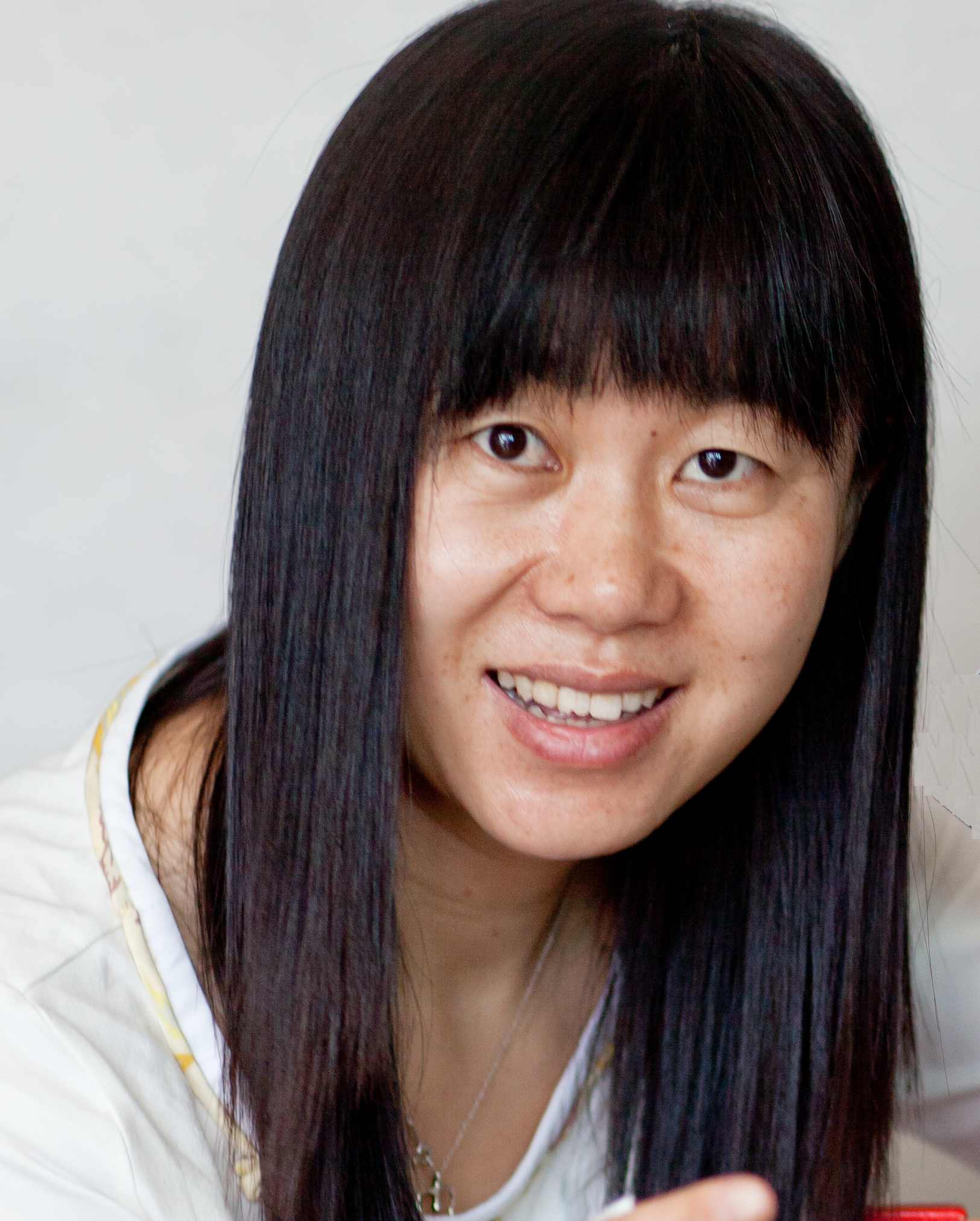}}]{Shixia Liu}
is an associate professor at Tsinghua University. Her research interests include visual text analytics, visual social analytics, interactive machine learning, and text mining. She worked as a research staff member at IBM China Research Lab and a lead researcher at Microsoft Research Asia.
She received a B.S. and M.S. from Harbin Institute of Technology, a Ph.D. from Tsinghua University.
She is an associate editor-in-chief of IEEE Trans. Vis. Comput. Graph.
\end{IEEEbiography}

\end{document}

%% file: introduction.tex

\section{Introduction}

\maketitle


Deep neural networks (DNNs) have 
demonstrated superior performance in many artificial intelligence applications, such as pattern recognition and natural language processing~\cite{Lecun2015_Survey,jiang2019recent,liu2017VI_towards}.
However, researchers have recently found that \jw{even} a 
\jw{highly accurate DNN can be} vulnerable to carefully-crafted adversarial \xia{examples that} are intentionally designed to mislead a DNN into making incorrect predictions~\cite{choo2018visual,Dezfooli2017_CVPR_Universal, Nguyen2015_CVPR_Deep,szegedy2013intriguing,Zheng2016_CVPR}.
For example, 
\jw{an attacker can make imperceptible modifications
\newdoc{
to} a panda image (from \pictag{I}{1} to \pictag{I}{2} in Fig.~\ref{fig:teaser-overview}) to mislead a state-of-the-art DNN model~\cite{He2016_CVPR_residual} to classify it as a monkey.}
This phenomenon
\newdoc{
creates} high risk
\newdoc{
when} applying DNNs to safety- and security-critical applications, such as driverless cars, face recognition ATMs, and Face ID security on mobile phones~\cite{Akhtar2018_threat}.
\mcrev{For example, researchers have recently shown that even the state-of-the-art public Face ID system can be fooled by using a carefully-crafted sticker on 
\newdoc{
a} hat~\cite{komkov2019advhat}.}
Thus, there is \jw{an urgent} 
need to understand the prediction process of adversarial examples
and identify the root cause of incorrect predictions~\cite{Akhtar2018_threat, wang2018interpret}.
Such an understanding is valuable for developing adversarially robust solutions~\cite{Goodfellow2015_ICLR_explaining,liao2018defense,pang2018towards}.
\newdoc{
A}
recent survey identifies two important
\jw{questions that require analysis}~\cite{Akhtar2018_threat}:
(1) why similar images (e.g., adversarial and normal panda images) \textbf{\normalsize diverge} \jw{into} 
different predictions, and 
(2) why images from different classes (e.g., adversarial panda images and normal monkey images) \textbf{\normalsize merge} \jw{into the} 
same prediction.


\jw{To give analytical answers to the questions,}
\keleiReRe{we need to solve two technical challenges.}
The first 
is \mc{\jw{to disclose} 
the prediction process of a DNN.
To this end, we need to extract the critical neurons and their connections that are responsible for the predictions of examples (Fig.~\ref{fig:system} (b)).
Such neurons and their connections form the datapaths of examples~\cite{wang2018interpret}.}
However, in a DNN, the neurons have complex interactions with each other~\cite{Bengio2013_PAMI_Representation}.
Thus, it is technically demanding to disentangle the roles of these neurons
\newdoc{
within the entire} network and \shixia{extract the critical neurons to} form the datapath.
The second challenge is to effectively illustrate \mc{and compare the prediction processes of adversarial and normal examples based on the extracted datapaths.}
A state-of-the-art DNN usually contains hundreds of layers, with millions of neurons in each layer~\cite{He2016_CVPR_residual}.
Thus, an extracted datapath potentially contains millions of neurons and even more connections.
Directly visualizing all the neurons and 
connections in \mc{the extracted datapath} will \jw{result in} 
excessive visual clutter.\looseness=-1

\mc{To tackle these challenges, we have developed a visual analysis tool, AEVis, to \jw{help identify} the root cause of \jw{misclassification of} adversarial examples.
Fig.~\ref{fig:teaser-overview} shows an example of using AEVis to analyze why an adversarial panda image is misclassified. 
On the one hand, we find that the extracted datapaths of the adversarial and normal panda images start to diverge at layer \kelei{\pictag{L}{A} (Fig.~\ref{fig:teaser-overview})} and eventually lead to 
different predictions.
On the other hand, merging starts at layer \kelei{\pictag{L}{C} (Fig.~\ref{fig:teaser-overview})} in the datapaths of 
the adversarial panda \jw{and monkey images}. 
With the \jw{use of the developed} 
multi-level visualization, we identify the root cause of this misclassification as 
\jw{both a failed detection of the outline of \newdoc{one of} the panda's
\newdoc{
ears} and a faulty detection of a monkey face in the adversarial panda image using the target DNN.}
}\looseness=-1

\mc{Technically, AEVis aims to disclose the prediction process of a DNN by extracting and visualizing \jw{the} datapaths for adversarial and normal examples, especially focusing on illustrating how \jw{these} datapaths \jw{diverge and merge.} 
}

\mc{To achieve 
\newdoc{
this} aim, 
we first formulate the datapath extraction as a subset selection problem, which aims
\newdoc{
to select} a minimum set of neurons that can maintain the predictions of a set of examples. 
As neurons in a DNN sometimes have similar roles, 
\jw{there is randomness in selecting neurons in the datapath extraction process.} 
\jw{As a result, the uniqueness of an example's extracted datapath cannot be guaranteed.}  
\jw{Moreover,} the randomness hinders the detection of the diverging and merging patterns \jw{in} 
the extracted datapaths.
To reduce the randomness, we 
\jw{introduce the} 
constraint that 
\newdoc{
it is desirable for the} datapaths of adversarial and normal examples
\newdoc{
}to share common feature maps (\todo{a set of neurons that share the same weights in a DNN}).   
To extract the datapath\newdoc{s} for large DNNs, we approximate the subset selection problem as a continuous optimization 
\jw{that can be} efficiently solved by constructing and training a neural network~\cite{wang2018interpret}.}

\mc{Second, we have developed a multi-level visualization that illustrates how the extracted datapaths diverge and merge in the prediction process.
In particular, at the network-level, we 
\newdoc{
have created} a river-based visualization to provide an overview of the diverging and merging patterns of datapaths.
At a detected diverging/merging point (layer level), we employ a treemap-based set visualization to  
\jw{illustrate the neuron groups at this layer and their
belonging to different datapaths.} 
\jw{This} 
helps experts determine the critical \jw{neurons} 
that cause 
the diverging and merging patterns.} 
\mc{In addition, we
\newdoc{
have enhanced} the multi-level visualization with a set of rich interactions that 
enable experts to effectively analyze \jw{the cause of diverging/merging of datapaths.} 
For example, we allow experts to interactively analyze the contribution of neurons in one layer to those of another deeper layer in order to disclose the root cause of a diverging/merging \jw{pattern} 
in the compared datapaths.}

The paper 
\newdoc{
is an extension of} our previous work~\cite{liu2018analyzing}, 
\newdoc{
in which} datapaths of examples are extracted and illustrated.
In this paper, we
\newdoc{
address} the problem 
\newdoc{
of} merging patterns
\newdoc{
that were} not detected in
\newdoc{
our} previous 
\newdoc{
method}.
We provide a better overview of diverging and merging between the datapaths of examples. 
In addition, the root cause of such patterns is analyzed
\newdoc{
more deeply} with our refined analysis workflow.
\mcrev{To evaluate the usefulness of the new system, we re-invited one expert in our previous work and conducted a deeper analysis of the same two adversarial image pairs, panda-monkey (Sec.~\ref{sec:panda}.1) and cannon-racket (Sec.~\ref{sec:cannon}.2).}
These improvements come from the following technical contributions:
\begin{compactitem}
\item\noindent{\textbf{\normalsize A constrained datapath extraction \jw{algorithm}
} to extract datapaths while preserving their diverging and merging patterns.
}
\item\noindent{\textbf{\normalsize A river-based visualization} to provide an overview of how datapaths diverge and merge at the network level and \textbf{\normalsize a refined layer-level visualization} to reveal the feature maps of interest.}
\item\noindent{\textbf{\normalsize A contribution analysis method} to iteratively investigate the contribution of neurons between two layers and help experts analyze the root cause of diverging/merging
\newdoc{
in} certain layers.} 
\end{compactitem}

\todo{In this paper, we focus on analyzing adversarial examples generated for convolutional neural networks (CNNs), because CNNs are among the most widely-used networks, and most of the current adversarial example generation methods focus on attacking CNNs~\cite{Akhtar2018_threat}.
Our method can \newdoc{also} be 
\newdoc{
used} to analyze adversarial examples
\newdoc{for} other deep networks that use CNNs as the key components.} 

%% file: related.tex
\section{Related Work}\label{sec:related-work}

In the field of visual analytics, a number of methods have been developed to illustrate the working mechanism \kelei{of} a variety of DNNs, such as CNN~\cite{pezzotti2018deepeyes, nie2018tnnvis, bilal2018convolutional}, RNN~\cite{ming2017understanding, strobelt2018lstmvis, kwon2018retainvis, strobelt2018seq2seq}, deep generative models~\cite{liu2017analyzing, wang2018ganviz, Kahng2018ganlab}, and deep reinforcement learning models~\cite{wang2018dqnviz}.
Hohman et al.~\cite{Hohman2018VisualAnalyticsSurvey} presented a comprehensive survey to summarize the state-of-the-art visual analysis methods for explainable deep learning.
Existing methods can be categorized into three classes:
network-centric~\cite{Liu2017_TVCG_Towards,Tzeng2005_VIS, wongsuphasawat2018visualizing}, instance-centric~\cite{rauber2017visualizing, bilal2018convolutional, wang2019deepvid, ma2019explaining}, and hybrid~\cite{Harley2015_ISVC,kahng2018acti}.\looseness=-1

\noindent\textbf{Network-centric methods.}
Network-centric methods help explore the
\newdoc{
entire} network structure of a DNN, illustrating the roles of neurons/neuron connections/layers in the training/test process.
\newdoc{
In the} pioneering work, Tzeng et al.~\cite{Tzeng2005_VIS} employed a DAG visualization to illustrate the neurons and their connections.
This method can illustrate the structure of a small neural network but suffers from severe visual clutter when visualizing state-of-the-art DNNs.
To solve this problem, Liu et al.~\cite{Liu2017_TVCG_Towards} developed a scalable visual analysis tool, CNNVis, based on clustering techniques. 
It helps explore the roles of neurons in a  
\jw{deep CNN} and diagnose 
failed training processes.
Wongsuphasawat et al.~\cite{wongsuphasawat2018visualizing} developed a tool with a scalable graph visualization to present the dataflow 
of a DNN.
To produce a legible graph visualization, they applied a set of graph transformations that converts the low-level \jw{graph of dataflow} 
to the high-level structure of a DNN.

The aforementioned methods 
\jw{help experts better understand} 
the network structure, but they are less capable of explaining the predictions of individual examples.

\noindent\textbf{Instance-centric methods.}
To address the aforementioned issue,
\keleiReRe{researchers made several recent attempts that
\newdoc{
focus on} instances.}
These attempts aim at analyzing the learning behavior of a DNN revealed by the instances.
\keleiReRe{A widely-used method is feeding a set of instances into a DNN and}
visualizing \shixia{the corresponding log} data, such as the 
activation or the final predictions.

For example, Rauber et al.~\cite{rauber2017visualizing} designed a compact visualization to reveal how the internal activation of training examples evolves during a training process.
They used t-SNE~\cite{maaten2008visualizing} to project the high-dimensional activation \jw{maps} of training examples in each snapshot 
\newdoc{
onto} a 2D plane.
The projected points are connected by 2D trails to provide an overview of the activation 
\newdoc{
during} the whole training process.
\mc{The method successfully demonstrated how different classes of instances are gradually distinguished by the target DNN.}
\newdoc{
In addition to} internal activation, the final predictions of instances can also help experts analyze the instance relationships.
For example, the tool Blocks~\cite{bilal2018convolutional} utilizes a confusion matrix to visualize the final predictions of a large number of instances.
\keleiReRe{To reduce the visual clutter caused by a large number of instances and classes, researchers enhanced the confusion matrix
\newdoc{
using} techniques such as}
non-linear color mapping and halo-based visual boosting.
\mc{The enhanced confusion matrix was able to disclose the confusion pattern among different classes of instances and further indicated the learning behavior of a target CNN.}

The above methods can provide an overview of a large number of instances and help experts analyze their relationships.
However, the prediction process of individual instances is less considered.
Compared with these macro-level methods, our method focuses on the micro-level and targets \newdoc{
}the prediction processes of a set of instances (usually a few to dozens).
The prediction processes of these instances are visualized 
\newdoc{
using} a multi-level datapath visualization.
Revealing the prediction processes enables experts 
\jw{to analyze} the root cause of \newdoc{the} 
\jw{misclassification of adversarial examples.}\looseness=-1

\noindent\textbf{Hybrid methods.} The hybrid methods combine the advantages of network-centric and instance-centric methods.
Like instance-centric methods, the hybrid methods also feed the target instances into the network and extract log data such as activation \jw{maps}.
The extracted log data is often visualized in the context of the network structure, which provides visual hints to select and explore the data of interest, e.g., the activation in a specific layer.
Visualizing the log data in the context of network structure also helps experts explore the data flow from the network input to the output~\cite{liu2018steering}.

There are several 
\newdoc{
papers making progress} in this direction.
For example, Hartley et al.~\cite{Harley2015_ISVC} developed an interactive node-link visualization to show the activation in a DNN.
Although this method is able to illustrate detailed activation on feature maps, it suffers from severe visual clutter when dealing with large CNNs.
To solve this problem, Kahng et al.~\cite{kahng2018acti} developed ActiVis to interpret large-scale DNNs and their results.
They employed a multiple coordinated visualization to facilitate experts in comparing activation among examples.
The above works mainly focus on exploring the prediction process of \textbf{normal} examples.
\todo{Recently, there is an emerging need 
\newdoc{
in} safety-critical fields to analyze \textbf{adversarial} examples of DNNs.
While machine learning researchers have developed some holistic views on understanding the existence of adversarial examples~\cite{Goodfellow2015_ICLR_explaining, liao2018defense}, 
\newdoc{
there is still a lack of} visualization tools to analyze the details.}
\newdoc{
In response to this} need, we developed AEVis~\cite{liu2018analyzing} to analyze the root cause of misclassifications produced by malicious \textbf{adversarial} examples.
In particular, we developed a datapath extraction \jw{method} 
to extract critical neurons and their connections in the prediction process.
To enable experts \jw{to} explore the extracted datapaths, we designed a multi-level visualization that presented datapaths from the high-level network structure to the detailed neuron activation.

As an extension of our previous work~\cite{liu2018analyzing}, this paper 
\jw{re-identifies} 
the central analytical task as analyzing the diverging and merging patterns of normal and adversarial examples.
Based on this task, we \jw{developed} 
a constrained datapath extraction method that better \jw{preserves} 
the diverging and merging patterns of \newdoc{
}normal and adversarial examples.
We also 
\newdoc{
enhanced}
the whole analysis workflow by \jw{introducing} several useful interactions, such as \todo{activation analysis and contribution analysis}.
These interactions enable the experts to gradually investigate the major reason
\newdoc{
for}
this diverging/merging pattern and thus help them analyze the 
\jw{misclassification of} 
adversarial examples.

%% file: system.tex
\section{The Design of AEVis}\label{sec:system}

\begin{figure*}[t]
  \centering
  \vspace{5mm}
  \begin{overpic}[width=0.95\linewidth]{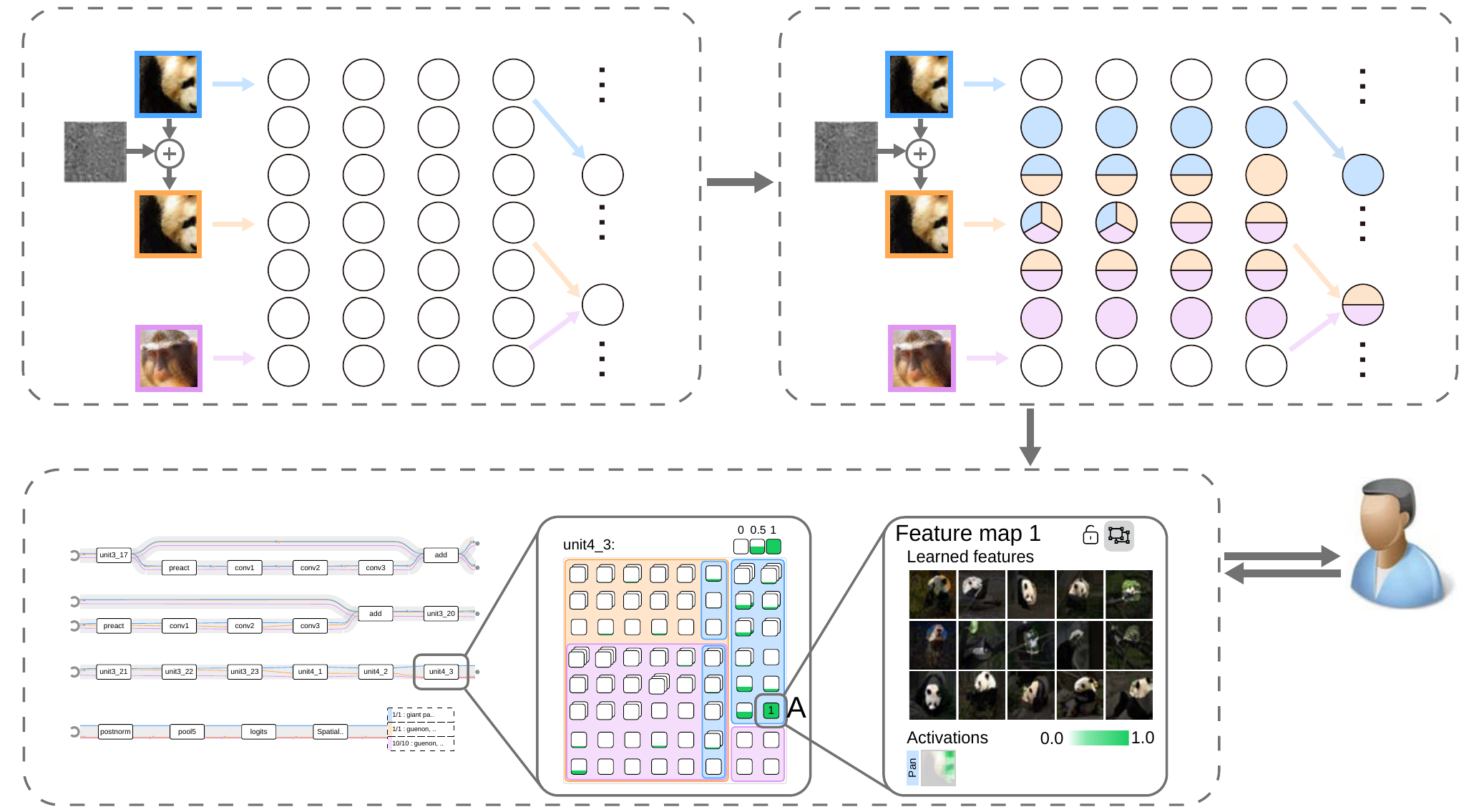}
  \scriptsize
  \put(45.2, 55.2){\textsf{(a)}}
  \put(12, 55){\textsf{Input Images and the Target Model}}
  \put(25.5, 51.7){\textsf{DNN}}
  \put(38, 51.7){\textsf{Prediction}}
  \put(39, 45.4){\textsf{Panda}}
  \put(38.5, 36.4){\textsf{Monkey}}
  \put(4.5, 52){\textsf{Normal source image}}
  \put(4.5, 47.2){\textsf{Noise}}
  \put(4.5, 36){\textsf{Adversarial image}}
  \put(4.5, 33.5){\textsf{Normal target image}}
  \put(96, 55.2){\textsf{(b)}}
  \put(70, 55){\textsf{Datapath Extraction}}
  \put(76.5, 51.7){\textsf{DNN}}
  \put(89, 51.7){\textsf{Prediction}}
  \put(90.5, 45.4){\textsf{Panda}}
  \put(90, 36.4){\textsf{Monkey}}
  \put(55.8, 52){\textsf{Normal source image}}
  \put(55.2, 47.2){\textsf{Noise}}
  \put(55.8, 36){\textsf{Adversarial image}}
  \put(55.8, 33.5){\textsf{Normal target image}}
  \put(80, 24){\textsf{(c)}}
  \put(34, 23.8){\textsf{Datapath Visualization}}
  \put(13, 20.7){\textsf{Network-level}}
  \put(42, 20.7){\textsf{Layer-level}}
  \put(65, 20.7){\textsf{Neuron-level}}
  \put(84.1, 13.8){\textsf{Feedback}}
  \end{overpic}
  \vspace{-2mm}
  \caption{\jw{AEVis system overview.} (a) Input of the AEVis system; (b) the datapath extraction module; (c) the datapath visualization module that illustrates the extracted datapaths \jw{at the network-, layer-, and neuron-level.}}
  \label{fig:system}
  \vspace{-2mm}
\end{figure*}

The development of AEVis
\newdoc{
was in collaboration} with the machine learning team that won \newdoc{
}first place in the NIPS 2017 non-targeted adversarial attack and targeted adversarial attack competitions, which aim\newdoc{ed} at attacking CNNs~\cite{dong2018boosting,NIPS2017_Challenge}.
Despite the promising results \jw{they achieved}, the experts found 
\jw{the research process} inefficient and inconvenient, especially \jw{in terms of} the explanation of the model outputs.
In their research process, a \jw{key} 
step 
\newdoc{
was} \jw{to explain the misclassification} 
introduced by adversarial examples.
Understanding why an error has been made \jw{helps the experts identify the model weakness} 
and further design a more effective 
\newdoc{
attack/defense} method.
The experts thus desire a tool that can assist them 
in understanding the prediction process of the target CNN.

\subsection{Requirement Analysis}
\label{sec:require}


We have identified the following high-level requirements based on previous research and 
discussions with two experts (E$_1$ and E$_2$) from the winning team of the NIPS 2017 competition.

\noindent \textbf{\normalsize R1 - \mc{Extracting the datapaths for adversarial and normal examples}.}
Both experts expressed the need for extracting the datapaths of \mc{adversarial} examples, \mc{which can disclose the prediction process of 
adversarial examples and thus} serves as the basis for analyzing why the adversarial examples 
\newdoc{
were} misclassified.
In a CNN, different neurons learn to detect different features~\cite{Zeiler2014_ECCV}, 
\jw{and play different roles} 
for the prediction of an example.
 E$_1$ said that only analyzing the datapath can greatly 
 \newdoc{
 reduce their effort} by allowing them to only focus on critical neurons \jw{rather than \newdoc{having} to examine all of them.} 
In addition to the datapaths for \mc{adversarial} examples, E$_1$ emphasized the need for extracting \mc{datapaths} for normal examples simultaneously.
\mc{He commented that as an adversarial example is often generated by slightly perturbing the pixel values of a normal image, \jw{there must be similarities between the two extracted datapaths.
Considering the similarity during the datapath extraction process will
\keleiReRe{help extract}
more meaningful datapaths for comparison \newdoc{
}during the analysis.}
}

\noindent \textbf{\normalsize \mc{R2 - Comparing the datapaths of adversarial and normal examples.}}
\mc{As mentioned before, an adversarial example is often \jw{generated} 
by adding unperceivable noise to a normal example,} 
\jw{and thus
\newdoc{
there is} little difference from the normal image in the input space.} 
However, their prediction results are different.
The experts are interested in how they diverge \newdoc{in}to different predictions.
For example, E$_2$ commented, ``I want to know whether there are some critical `diverging points' for the \jw{different predictions} 
or 
\newdoc{
they accumulate} gradually layer by layer through the network.''
To this end, E$_2$
\newdoc{
wanted} to compare the datapaths of normal source examples and adversarial examples.
Triggered by E$_2$, E$_1$ added that it was interesting to compare the datapath of an adversarial example (e.g., a panda image that is misclassified as a monkey) with that of normal target examples (e.g., normal monkey images). 
Such comparisons help understand how these very different images ``merge'' into the same prediction (e.g., the monkey). 
The need for visual comparison is consistent with the findings of previous research~\cite{alexander2016task, gleicher2018considerations,liu2018visual}.\looseness=-1


\noindent \textbf{\normalsize R3 - Exploring datapaths at different levels.}
In a large CNN, a datapath often contains millions of neurons and connections.
Directly presenting all neurons in a datapath will induce severe visual clutter.
E$_1$ commented, ``I cannot examine all the neurons in a datapath because there are too many of them.
\jw{Instead,} I often start by selecting an important layer based on my knowledge and examine the neurons in that layer to analyze the learned features and the activation of these neurons.
The problem 
is \newdoc{that} when dealing with a new architecture, I may not know which layer to start with.
Thus, I have to examine a bunch of layers, which is very tedious.''
\jw{He advocated \newdoc{for} the idea}
\newdoc{
of providing} 
an overview of the datapath with visual guidance to facilitate\newdoc{
} experts in selecting the layer of interest.
The requirement of providing an overview of a CNN aligns well with previous research~\cite{kahng2018acti,liu2018analyzing,wongsuphasawat2018visualizing}.
Although the overview of a datapath facilitates experts in finding the layer of interest, it is not enough to diagnose the root cause of the wrong prediction.
The experts said that 
\jw{a 
\newdoc{
link} between} 
the overview of a datapath and the 
detailed neuron activation \jw{is required, which} 
helps them identify the most important neurons that lead to misclassification.
\jw{To summarize,} 
it is
\newdoc{
desirable} to provide a multi-level exploration mechanism that allows experts to zoom into the neurons of interest gradually.
Previous research also indicates that visual analytics for deep learning benefits from multi-level visualization~\cite{kahng2018acti,liu2018analyzing}.

\noindent \textbf{\normalsize R4 - \mc{Examining how neurons contribute to each other in a datapath}.}
Finding a diverging or merging point is not the end of the analysis.
To develop effective defense methods,
\keleiReRe{we must}
disclose how such divergence or
\newdoc{
merging} happens.
As the data flows from previous layers to the current diverging or merging point, a practical method of finding the root cause is tracing back to the previous layers and examining how \newdoc{the} neurons \jw{there} contribute to the neurons
\newdoc{
at} the diverging or merging point.
E$_1$ commented, ``When I find a neuron or feature map that performs very differently for an adversarial and a normal example, I'm interested in \jw{the cause of} 
\newdoc{
this} difference. 
For example, it is useful to know whether it \newdoc{was} caused by the neurons in the previous layer or even the neurons in a far-away layer due to the skip-connections~\cite{He2016_CVPR_residual} in modern CNNs.''
\jw{Therefore,} we need to analyze how neurons contribute to each other in a DNN.
Previous research also indicates that presenting the contribution\newdoc{s} among neurons is \jw{important} 
for understanding the outputs and roles of neurons~\cite{liu2018analyzing}.

\subsection{System Overview}

Driven by the requirements
\newdoc{
suggested by these} experts, we have developed a visual analysis tool, AEVis, 
\mc{to help experts analyze the root cause of \jw{the 
robustness issues arising from adversarial examples.}}
It consists of the following two parts. 
\begin{compactitem}
\item\noindent{\textbf{\normalsize A datapath extraction module} that extracts the critical neurons and their connections for \mc{the predictions of adversarial and normal examples} (\textbf{R1}).}
\item\noindent{\textbf{\normalsize A datapath visualization module} that \mc{enables a multi-level (\textbf{R3}) visual comparison (\textbf{R2}) of the extracted datapaths and provides rich interactions (\textbf{R4}) to analyze the root cause of a misclassification}.} 


%
\end{compactitem}

As shown in Fig.~\ref{fig:system} \todo{(a)}, AEVis takes a trained CNN and the examples to be analyzed as
\newdoc{
its} input.
The examples \mc{usually} include \mc{
the adversarial examples, normal source examples, and normal target examples.}
Given the examples and the CNN, the datapath extraction module extracts the critical \mc{neurons} and 
their connections that are responsible for the predictions of the examples (Fig.~\ref{fig:system} (b)).
The extracted datapaths are then fed into the visualization module (Fig.~\ref{fig:system} (c)), which supports the navigation and comparison of the datapaths from the high-level layers to the detailed neuron activation.\looseness=-1

%% file: datapath.tex
\section{Datapath Extraction}
\label{sec:datapath}


\subsection{Basic Problem Formulation}

\mc{Extracting datapaths of adversarial and normal examples is the basis for analyzing why an adversarial example is 
misclassified (\textbf{R1}).
The key challenge is to identify the critical neurons in the prediction process. \jw{Once the critical neurons 
\newdoc{
have been} identified, selecting the corresponding connections to form the datapath is straightforward.} 
}
Critical neurons are \jw{those} 
that highly contribute to the final prediction.
In other words, by only combining \jw{the critical neurons and corresponding connections,} 
the prediction of an example will not be changed. 
\jw{Therefore,} we aim 
\newdoc{
to select} a minimized subset of neurons \jw{that can} 
\newdoc{
maintain} the original prediction. 
Accordingly, we formulate critical neurons extraction as a subset selection problem:
\begin{equation}
\label{eq:dp-ext2}
N^{opt}=\underset{{N_s}\subseteq N}{\mathop{\arg \min }}\,(p(x)-p(x;{N_s}))^2 +\lambda|N_s|.
\end{equation}
The first term is to keep the original prediction, and the second term ensures 
\newdoc{
the selection of} a minimized subset of neurons.
Specifically, $N$ is the set of neurons in a CNN, $N_s$ is a subset of  
$N$, $N^{opt}$ is the \jw{optimized subset consisting of} 
critical neurons, $p(x)$ is the prediction of example $x$, and $p(x;N_s)$ is the prediction if we only consider the neuron subset $N_s$.
To measure the difference between two predictions, we adopt the widely used $\ell_2$ norm.
$|N_s|$ is the size of $N_s$ and $\lambda$ is used to balance the two terms.
\mc{
Compared with our previous work, we change the second term from $|N_s|^2$ to $|N_s|$.
With
\newdoc{
this} change, we are able to accelerate the
\newdoc{
entire} optimization process by obtaining a minimized subset of neurons more easily according to the Lasso algorithm~\cite{tibshirani1996regression}.}

The large search space in Eq.~(\ref{eq:dp-ext2}) hinders \jw{a direct solution,} 
which is mainly due to a large number of neurons in a CNN (usually millions).
To reduce the search space, we utilize the weight-sharing property in CNNs~\cite{Lecun2015_Survey} and group neurons into a set of feature maps.
Specifically, in a CNN, neurons in a feature map share the same weights, and thus learn to detect the same feature.
\jw{Making use of}
\newdoc{
this} characteristic, we 
\jw{replace the problem of critical neuron selection with feature map selection and reformulate the problem as:} \looseness=-1
\begin{equation}
\label{eq:dp-ext-fm}
F_{opt}=\underset{{F_s}\subseteq F}{\mathop{\arg \min }}\,(p(x)-p(x;F_s))^2 +\lambda|F_s|,
\end{equation}
where $F$ is the set of feature maps in a CNN, \newdoc{and} $F_s$ is a subset of $F$.

\begin{figure}[bpht]
  \centering
  \begin{overpic}[width=0.97\linewidth]{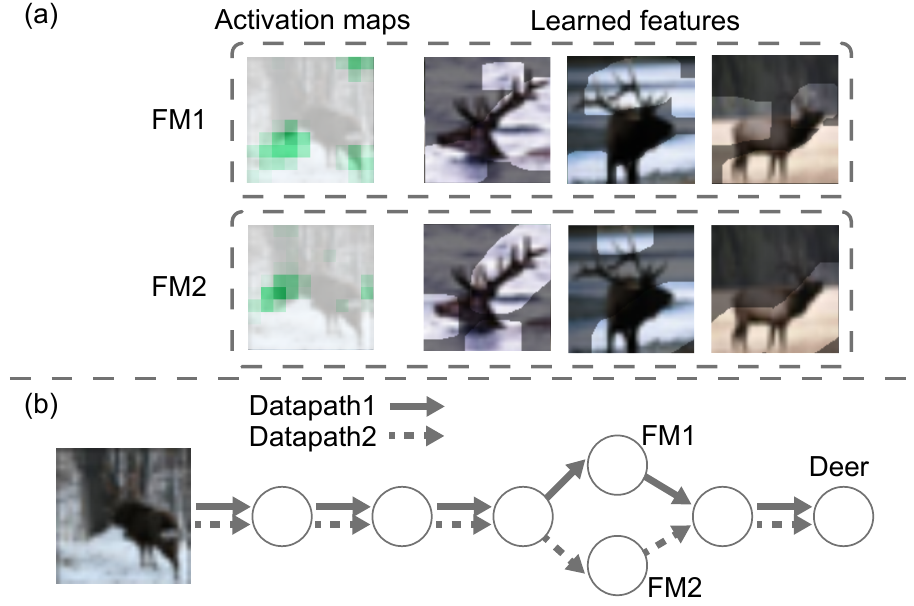}
  \end{overpic}
  \caption{
  The cause of randomness in datapath extraction. (a) two feature maps detect the same feature (a deer head); (b) there are two 
  equivalent candidate datapaths 
  for the deer image.
  }
  \label{fig:expr_randomness}
\end{figure}

\begin{figure*}[!t]
  \centering
  \begin{overpic}[width=\linewidth]{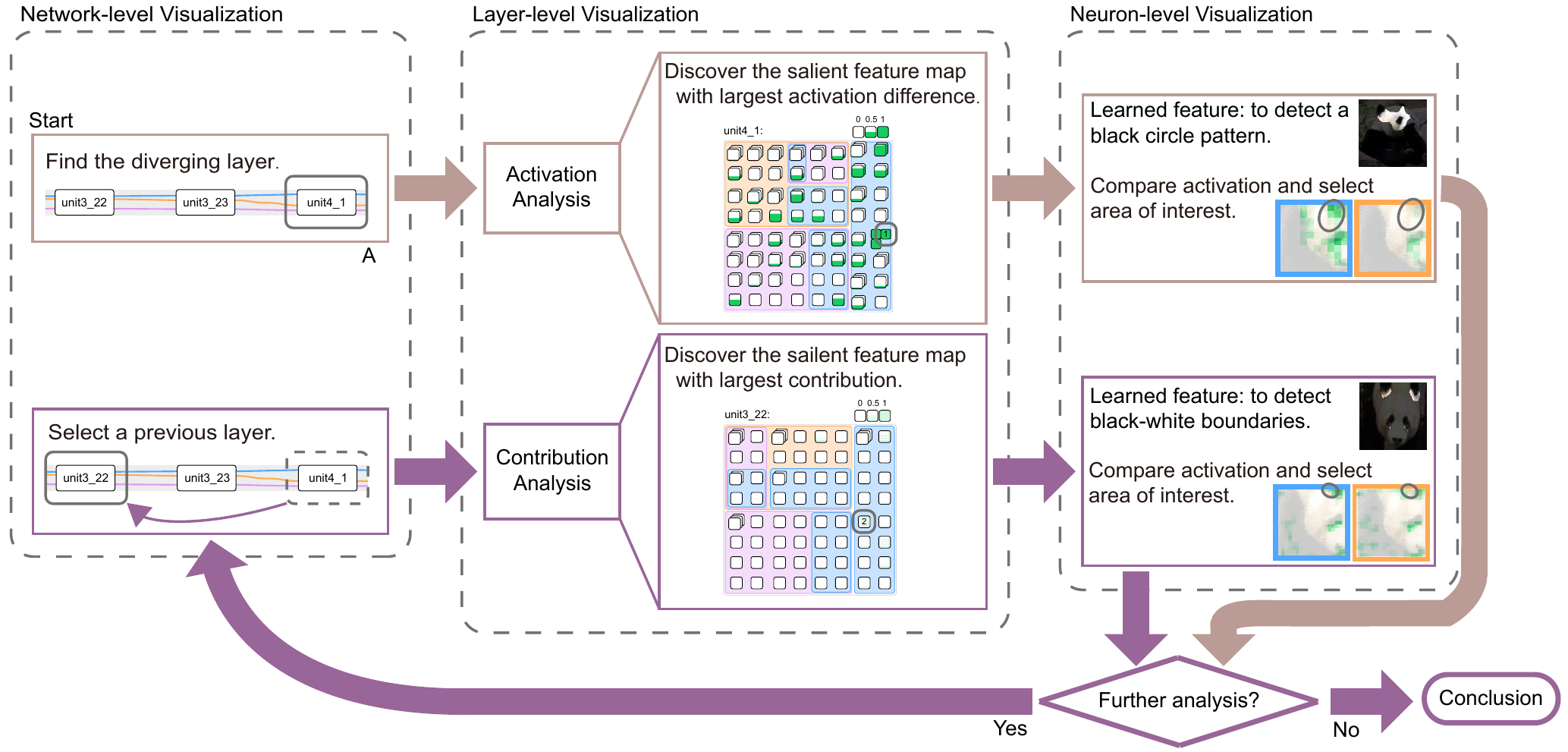}
  \end{overpic}
  \caption{The analysis workflow of the diverging pattern. The brown color represents the first step analysis, and the purple color indicates the subsequent analysis, which is iterative.}
  \label{fig:analysis_flow}
\end{figure*}

\subsection{Constrained Datapath Extraction}

\mc{The above method is successful in extracting the critical feature maps for \textbf{one} example but sometimes
\newdoc{
creates difficulty when} \textbf{comparing} datapaths of adversarial and normal examples, especially in 
\newdoc{
the detection of} merging patterns~\cite{liu2018analyzing}.
After
\newdoc{
discussions} with the domain experts (E$_1$ and E$_2$) and conducting several experiments,
we find that the difficulty is mainly due to the randomness in datapath extraction.
Specifically, different feature maps in a CNN may have \jw{very similar} 
roles, i.e., detecting nearly the same features~\cite{liu2017VI_towards}.
It means that \jw{the optimized datapath for an individual adversarial or normal example may not be unique, given the many feature maps with equivalent roles.}}
\mc{
Thus, extracting a datapath can be treated as sampling one from \newdoc{
}equivalently good candidate datapaths, which introduces randomness in\newdoc{to} the datapath extraction.
Extracting datapaths that share common feature maps may lead to two feature map subsets
\newdoc{
that lack} common feature maps.   
As a result, we may over-estimate the difference between two extracted datapaths, which hinders the detection of the diverging and especially the merging patterns.}

\mc{To illustrate the above analysis, we trained a 6-layer CNN on the CIFAR10 dataset~\cite{Krizhevsky2009_Report}.
The network contains 5 convolutional layers and 1 fully connected layer.
After training, two equivalently good datapaths for a deer image (the difference between values of Eq.~(\ref{eq:dp-ext-fm}) is less than \kelei{$0.127$) \newdoc{
were} extracted}.
By examining the feature maps in the two datapaths, we found two feature maps that detected the same feature (a deer head, Fig.~\ref{fig:expr_randomness} (a)) but 
belong\newdoc{ed} to \newdoc{
}different datapaths (Fig.~\ref{fig:expr_randomness} (b)).
The above experiment illustrates that feature maps in a CNN may detect
the same feature and thus perform similar roles.
As a result, there is randomness in the datapath extraction process.
This randomness hinders the detection of diverging and merging patterns when comparing datapaths.}\looseness=-1

To faithfully disclose the diverging and merging patterns,
\keleiReRe{we need to reduce}
the randomness in \jw{datapath extraction.}  
The randomness is mainly caused by
\newdoc{
the lack of} preference among the feature maps that detect the same features.
To tackle this issue, we introduce additional constraints into the datapath extraction process, to
\newdoc{
prioritize the extraction of} datapaths that share common feature maps.

\mc{
Accordingly, the datapaths $F_{opt}^1, ..., F_{opt}^i,..., F_{opt}^n$ for examples $x^1,...,x^i,...,x^n$ are extracted by optimizing:
\begin{equation}
\label{eq:dp-ext-new-1}
F_{opt}^1, ..., F_{opt}^i,..., F_{opt}^n = \underset{{F_s^i}\subseteq F}{\mathop{\arg \min }}\,\sum_{i}{L^i}+
\gamma \sum_{i,j}{dis(F_s^i,\ F_s^j)},
\end{equation}
where the first term $L^i = (p(x^i)-p(x^i;{F_s^i}))^2 + \lambda |F_s^i|$ measures how good the datapath is for the i-th example $x^i$.
The second term $dis(F_s^i,\ F_s^j)$ \jw{is the distance to measure the difference between the datapaths for the i-th and j-th example, which is defined in Eq.~(\ref{eq:dp-ext-new-3}).}
Adding the constraint helps extract datapaths that share common feature maps.
$\gamma$ is used to balance the two terms.
Although this method is theoretically sound, in practice, we find 
that it
\newdoc{
is} difficult to maintain all the predicted labels of the examples, \jw{a fundamental requirement for explaining the prediction process.}
The root cause of the problem is the complexity in jointly finding optimized datapaths for different examples.
To solve this problem, we instead approximate the joint optimization into a chain of simpler conditional optimizations.
We first obtain the datapath for one example (e.g., the adversarial example to analyze) and iteratively obtain others by treating the previously calculated datapaths as constraints.
In particular, for the $i$-th example, we solve:
\begin{equation}
\label{eq:dp-ext-new-2}
F_{opt}^i= \underset{{F_s^i}\subseteq F}{\mathop{\arg \min }}\,L_i+
\gamma \sum_{j=1}^{i-1}{dis(F_s^i, F_{opt}^j)}.
\end{equation}}

\mc{To efficiently solve the above subset selection problem, we approximate this NP-hard discrete optimization~\cite{cormen2009introduction} with a continuous optimization:
\begin{equation}
\label{eq:dp-ext-new-3}
\begin{aligned}
& \vecz{opt}^i = \underset{\vecz{}^i\in [0,1]^n}{\mathop{\arg \min }}\, L(x^i, \vecz{}^i) + \gamma \sum_{j = 1}^{i-1} dis(\vecz{opt}^j, \vecz{}^i), \\
& L(x^i, \vecz{}^i)) = (p(x^i)-p(x^i;\vecz{}^i))^2 + \lambda |\vecz{}^i|, \\
& dis(\vecz{opt}^j, \vecz{}^i) = ||\vecz{opt}^j - \vecz{}^i||_2,
\end{aligned}
\end{equation}
where $\vecz{}^i=[z^i_1, \cdots , z^i_n]$ and $z_k^i \in [0,1]$ is the contribution of the $k$-th feature map in the datapath of the $i$-th example $x^i$.
We apply the commonly-used $\ell_2$ norm to measure the difference between two datapaths (the second term in Eq.~(\ref{eq:dp-ext-new-2})).
Eq.~(\ref{eq:dp-ext-new-3}) is further solved by constructing and training a DNN \jw{as in}~\cite{wang2018interpret}.
In particular, we embed the variable $\vecz{}^i$ into 
the target DNN and train the network on the target adversarial/normal examples by stochastic gradient descent (SGD).\looseness=-1
}

%% file: visualization.tex
\section{Datapath Visualization}

\subsection{Overview}

An extracted datapath usually contains millions of neurons and even more connections,
which prohibits 
efficient \jw{examination of the datapath or discovery of the merging-diverging patterns.} 
To help experts systematically investigate the extracted datapaths, we \newdoc{have} design\newdoc{ed} a multi-level visualization to facilitate the datapath analysis from the high-level network structure to the detailed neuron activation (\textbf{R3}).
Accordingly, it consists of three major visualization components at the network-, layer-, and neuron-levels.




\noindent \textbf{\normalsize Network-level visualization of data flows.}
As shown in Fig.~\ref{fig:system} (c), the network-level visualization provides an overview of the extracted datapaths, discloses the potential diverging and merging points, and further guides experts in selecting a 
\jw{layer of interest for examination} 
(\textbf{R2}).
\mcrev{Compared with our previous work, we replace 
the dot-plot-based network-level visualization with 
a river-based visual metaphor, \keleiReRe{which has
better scalability} and is more effective in depicting \newdoc{
} diverging and merging patterns.}
\noindent \textbf{\normalsize Layer-level visualization of feature maps.}
When an expert identifies a layer of interest (e.g., a diverging or merging point), s/he then zooms in 
to examine the critical feature maps in that layer (Fig.~\ref{fig:system} (c)).
For a diverging point, the unique feature maps of each datapath lie in the center of experts' analysis.
\newdoc{
For} a merging point, the shared feature maps between/among datapaths are critical
\newdoc{
to} the analysis.
To help experts \keleiRe{
more} quickly find the important and informative feature maps,
\keleiReRe{we use two types of filling styles}
to encode the activation \jw{difference} (solid filling $\vcenter{\hbox{\includegraphics[height=1.5\fontcharht\font`\B]{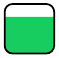}}}$, Fig.~\ref{fig:system}A) and contribution (dotted filling $\vcenter{\hbox{\includegraphics[height=1.5\fontcharht\font`\B]{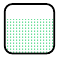}}}$, 
Fig.~\ref{fig:teaser-overview}$\textsf{F}_\textsf{A}$).
The higher filling 
represents a larger value.

\noindent \textbf{\normalsize Neuron-level visualization of learned features.}
When an expert finds a feature map of interest,
AEVis helps him/her understand what features the neurons of interest have learned in the prediction process.
Following previous research~\cite{bilal2018convolutional, liu2017VI_towards}, we employ the learned features of the neurons $\vcenter{\hbox{\includegraphics[height=1.5\fontcharht\font`\B]{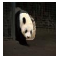}}}$ (Fig.~\ref{fig:teaser-overview}A) and their activation \jw{maps}  $\vcenter{\hbox{\includegraphics[height=1.5\fontcharht\font`\B]{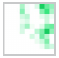}}}$ (Fig.~\ref{fig:teaser-overview}C) to facilitate the understanding.
The activation of a neuron in a feature map is encoded by the color.
Darker green indicates a higher value.

\noindent\textbf{Analysis workflows}.
These three visualizations work together to support a progressive analysis of adversarial examples, which helps
\newdoc{
experts} understand the root cause of the \newdoc{
divergence} between normal source examples and the corresponding adversarial examples, as well as the merging between the adversarial examples and the normal target examples. 
Fig.~\ref{fig:analysis_flow} shows the typical \jw{workflow} 
for analyzing a diverging pattern. 
It starts from the network-level visualization where a diverging pattern (Fig.~\ref{fig:analysis_flow}A) with several layer groups is identified first. Then \newdoc{
with} the layer-level visualization and activation analysis, the salient feature map is discovered.
Next, by analyzing the learned features and activation in the neuron-level visualization, the expert
\newdoc{
can identify} an area of interest in the focused feature map, which is sent to the contribution analysis module.
This module computes the contribution \jw{to the activation of the selected neurons} 
from corresponding neurons in previous feature maps. 
Finally, by examining the contribution of the feature maps \jw{in} 
the diverging pattern, the expert gradually investigates the major reason 
\newdoc{
for} this \jw{divergence.} 
In the merging pattern analysis, instead of using activation analysis, we use contribution analysis as the first step.
This is because the merging point is usually followed by the prediction. 
Contribution analysis helps identify \jw{the most important learned feature} 
for the final prediction.
\mcrev{}

With this exploratory analysis, the potential \jw{cause} 
for the wrong predictions is disclosed to facilitate experts in their task of noise robustness analysis. 
In the below sections, we focus on introducing the network-level visualization, the layer-level visualization, and contribution analysis.

\subsection{Network-level visualization}
\label{sec:vis-net}
\begin{wrapfigure}{r}{0.15\textwidth}
  \begin{overpic}[width=0.9\linewidth]{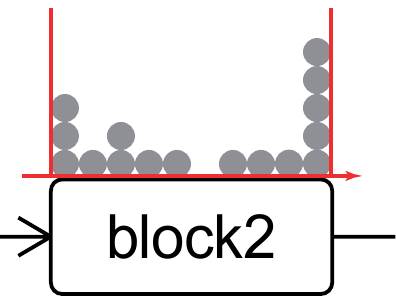}
    \put(15, 70){0.0}
    \put(65, 70){1.0}
  \end{overpic}
 \vspace{1mm} 
\caption{Dot plot. }
 \label{fig:dot}
 \end{wrapfigure}

In our previous work~\cite{liu2018analyzing}, we employed a dot plot to visualize the difference between \textbf{two} datapaths.
As shown in Fig.~\ref{fig:dot}, each rectangle represents a layer group, \mcrev{where layers are hierarchically grouped according to the hierarchical computation graph defined in the widely-used TensorFlow Graph Visualization~\cite{wongsuphasawat2018visualizing}.}
Each dot in the plot represents the activation similarity between two datapaths of a layer.
As a result, the dot plot is combined with the layer group to illustrate the similarities between the extracted datapaths of each layer in each layer group.
The position of a dot on the x-axis denotes the similarity value, from 0 (left) to 1 (right).
The method has been demonstrated to be useful in detecting the diverging/merging point of two datapaths (e.g., the datapaths of adversarial panda and normal panda images).
However, we have received feedback from the experts that the dot-plot-based visualization 
is less intuitive in revealing the overall evolution pattern of datapath merging and diverging as well as the transition between them. 
Moreover, it cannot compare three datapaths, which is specifically requested by the experts.
The experts said that in 
analysis, they often needed to examine the adversarial examples in the context of \jw{both} normal source examples (e.g., panda) and normal target examples (e.g., monkey) 
to identify the critical diverging/merging points.

\begin{figure}[bpht]
  \centering
  \begin{overpic}[width=0.95\linewidth]{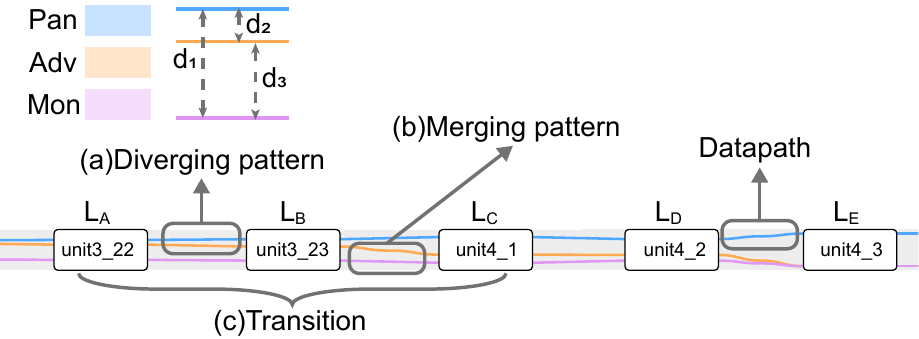}
  \end{overpic}
  \caption{
  \kelei{Visualization of three datapaths, with the illustration of (a) the diverging pattern, (b) the merging pattern, and (c) the transition from diverging to merging.}
  }
  \label{fig:case0_overview}
\end{figure}

To tackle these issues, we \newdoc{have} develop\newdoc{ed} a river-based visual metaphor~\cite{Cui2011textflow}, which is inspired by the natural phenomenon of \newdoc{a} river merging and diverging along
\newdoc{
a} riverbed.
The river-based visualization has been 
\newdoc{
proven} effective
\newdoc{
at} depicting
\newdoc{
}diverging and merging patterns over time~\cite{Cui2011textflow}.
As shown in Fig.~\ref{fig:case0_overview}, we use a curve to represent a datapath.
\mcrev{Considering the complexity of the current system, we do not use the curve width to encode extra information, and thus the width is always the same.}
The distance between \jw{the curves} represents the similarity between two datapaths.
The smaller the distance, the more similar the two datapaths.
\mc{\newdoc{
When} comparing three datapaths (adversarial, normal source, and normal target examples), we employ a rule-based method to highlight \newdoc{
}diverging and merging patterns.
In particular, the datapath of the adversarial example stays in the middle
\newdoc{
with} the other two (source and target) on
\newdoc{
either} side (Fig.~\ref{fig:case0_overview}).
The screen distance is proportional to the datapath distance $d_1$ (source-target).
The position of the datapath of the adversarial example is determined by
\newdoc{
retaining} the ratio of $d_2$ (adversarial - source) and $d_3$ (adversarial - target).}
To better reveal how data flows in the network, we embed the river-based visualization into the DAG visualization representing the network structure (Fig.~\ref{fig:case0_overview}). 
With this combination, 
the \textbf{merging} (Fig.~\ref{fig:case0_overview} (a)), \textbf{diverging}  (Fig.~\ref{fig:case0_overview} (b)), and 
\textbf{transition} between \jw{datapaths} 
can be easily recognized by examining the distance changes. 
(Fig.~\ref{fig:case0_overview} (c)).
For example, in Fig.~\ref{fig:case0_overview} (a), the distance between the blue curve (normal panda) and the orange curve (adversarial example)
\newdoc{
increases}.
This indicates that the critical neurons of these two datapaths are gradually
\newdoc{
becoming} less similar to each other, creating a \textbf{diverging pattern}.
While in Fig.~\ref{fig:case0_overview} (b), the distance between the orange curve (adversarial example) and the purple curve (normal monkey)
\newdoc{
decreases}.
This indicates that the critical neurons of these two datapaths are gradually 
\newdoc{
becoming} more similar to each other, creating a \textbf{merging pattern}.
Fig.~\ref{fig:case0_overview} (c) shows a transition process from the \jw{diverging} 
between a normal panda and an adversarial panda to the \jw{merging of} 
an adversarial panda and a normal monkey. 
Revealing these patterns helps experts quickly locate the layer of interest for further investigation.

\subsection{Layer-Level Visualization}
\begin{figure}[!t]
  \centering
  \vspace{4mm}
  \begin{overpic}[width=0.98\linewidth]{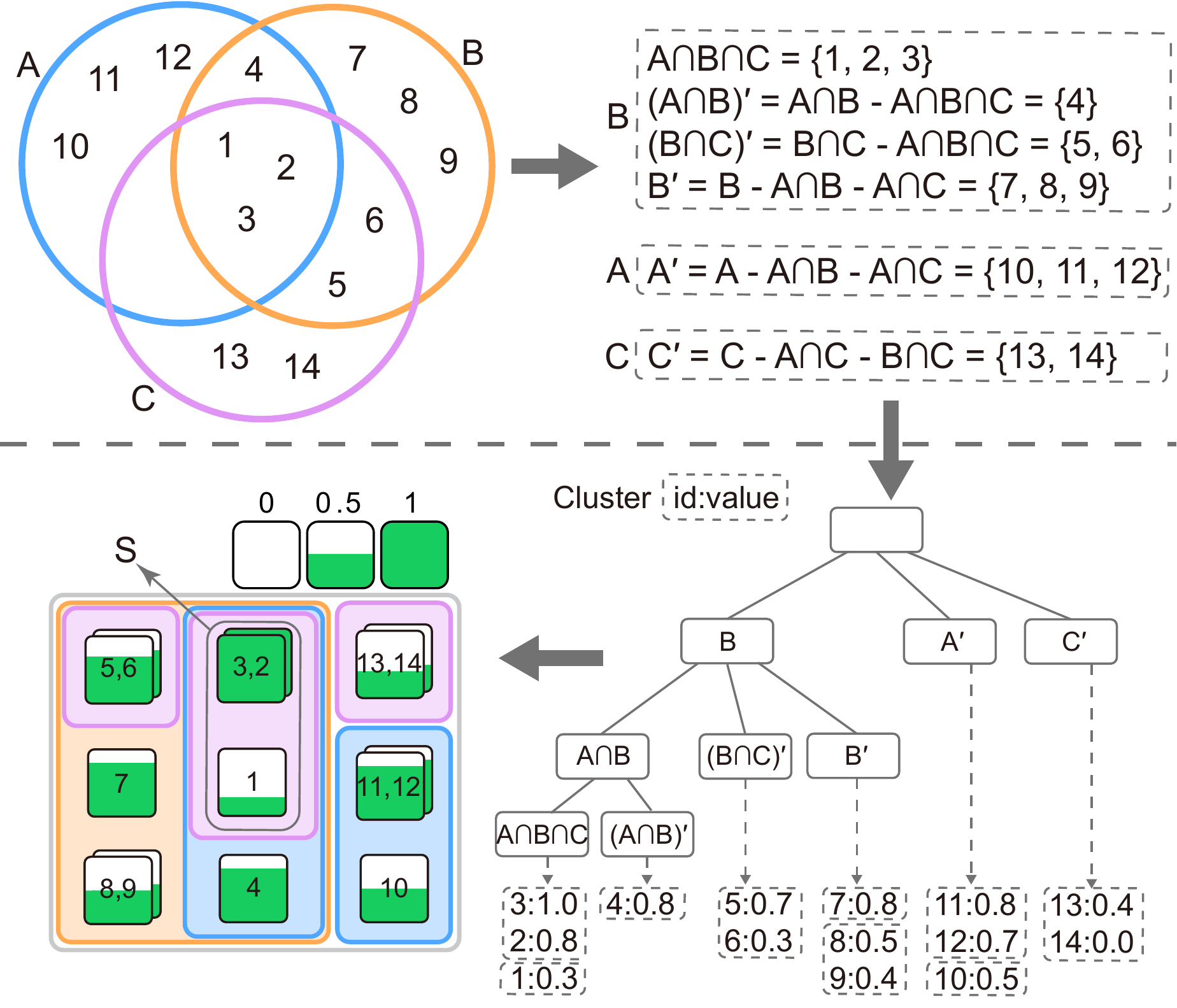}
  \small
  \put(20, 88){\textsf{(a)}}
  \put(70, 88){\textsf{(b)}}
  \put(20, -2){\textsf{(d)}}
  \put(70, -2){\textsf{(c)}}
  \end{overpic}
  \caption{
  Illustration on how to create a feature map visualization for three datapaths. 
(a) The three sets of feature maps 
at a selected layer;
(b) The intersection relations\newdoc{hips} among the sets;
(c) 
A hierarchy based on the set inclusion relations\newdoc{hips};
(d) The treemap-based visualization of feature maps. 
}
\label{fig:fm-layout-intro}
\end{figure}

When examining a layer of interest, such as a layer with a diverging or merging pattern, the critical feature maps of that layer (Fig.~\ref{fig:system} (c)) are \jw{important for understanding the key features learned in that layer.} 
\jw{These feature maps and corresponding learned features} are generally useful \jw{for understanding} 
why an adversarial example \jwReRe{diverges} 
from its original category and \jwReRe{merges} 
into another category.
As a result, the unique and shared feature maps between/among datapaths are expected to be encoded and visualized clearly.
To this end, we employ a treemap-based visualization to describe the set relations\newdoc{hips} among \jw{the feature maps} 
of different datapaths. 
The basic idea 
is illustrated in Fig.~\ref{fig:fm-layout-intro}.
\jw{In the figure,} 
(a) shows 
three \jw{sets of feature maps belonging to} 
three datapaths (normal panda, adversarial panda, and normal monkey), and their set relations.
We first compute the shared (intersection) and unique parts of \jw{the three sets} 
(Fig.~\ref{fig:fm-layout-intro} (b)). Then a hierarchy is built based on the set inclusion relations\newdoc{hips} (Fig.~\ref{fig:fm-layout-intro} (c)). 
To have more space for displaying the shared \newdoc{
}and unique part\newdoc{s} and distinguishing them clearly, we put the shared 
parts into the largest set with more feature maps.
Finally, the feature map sets and their intersection relations\newdoc{hips} are visualized with a squarified treemap layout~\cite{bruls2000squarified} (Fig.~\ref{fig:fm-layout-intro} (d)).

To better reveal the relationships 
between the shared and unique parts of different datapaths, the \jw{treemap cells of the shared parts} 
are placed \jw{
\newdoc{
in} a 
position
\newdoc{
that is as} close \newdoc{as possible} to every} 
treemap cell of 
related feature map sets. 
For example, the shared part, \textbf{S}, of feature map sets \textbf{A}, \textbf{B}, and \textbf{C} are placed near the centers of the three related treemap cells representing \textbf{A}, \textbf{B}, and \textbf{C} (Fig.~\ref{fig:fm-layout-intro} (d)). 
Accordingly, the treemap layout is formulated as an optimization problem 
with the goal of placing the \jw{treemap cells of} shared 
sets close 
to the center of \jw{the cells of related sets:} 
\begin{equation}
\begin{aligned}
\label{eq:treemap-optimization}
\min \sum_{s_i \in \{ 0,1\},\sum_i{s_i}\ge 2}
(f_e(\mathop{\cap}\limits_{i:s_i=1}{A_i}) -\mathop{f_m}\limits_{i:s_i=1}(f_e(A_i)))^2,
\end{aligned}
\end{equation}
where $\mathop{\cap}\limits_{i:s_i=1}{A_i}$ is the set that contains all feature maps shared by $A_i$, for all $s_i=1$. 
$s_i$ is the status variable of $A_i$. 
$s_i=1$ indicates that $A_i$ contains the shared part $\mathop{\cap}\limits_{i:s_i=1}{A_i}$, while $s_i=0$ means that $A_i$ does not contain this set.
$f_e(\cdot)$ denotes the center of the treemap cell representing a set, while $\mathop{f_m}\limits_{i:s_i=1}(\cdot)$ is the mean of the centers. 
Accordingly, the first term represents the center of the shared feature map set $\mathop{\cap}\limits_{i:s_i=1}{A_i}$, and the second term represents the mean of the centers of the feature map sets that share $\mathop{\cap}\limits_{i:s_i=1}{A_i}$.

This \jw{optimized} treemap-based visualization 
can clearly reveal the shared and unique feature maps \jw{on} 
the datapaths of interest, which is useful for investigating the roles of different types of feature maps (e.g., unique or shared feature maps) in the prediction.
For example, the experts are interested in examining the unique feature maps on each datapath for a diverging point.
While for a layer with a merging pattern, the shared feature maps among datapaths are critical for the prediction analysis.

To facilitate the identification of salient feature maps, two types of encoding are employed for activation and contribution, respectively.

\noindent\textbf{Encoding the activation difference}.
We select the maximum neuron activation in a feature map to represent its activation, with
\newdoc{
the aim of emphasizing} the most salient feature detected by the feature map.
The solid filling style $\vcenter{\hbox{\includegraphics[height=1.5\fontcharht\font`\B]{fig/glyph_solid.pdf}}}$ is used to encode the activation difference between two datapaths.
Taking datapaths \textbf{A} and \textbf{B}
\newdoc{
as an} example, their activation difference is
\textit{acti}(\textbf{A})-\textit{acti}(\textbf{B}). 
The larger \newdoc{the} value, the higher \newdoc{the} filling, 
\jw{which} indicates that the learned feature \jw{is more salient in \textbf{A} than in \textbf{B}.} 

\noindent\textbf{Encoding the contribution}.
\jw{A subset of neurons in a specific feature map can be selected as an area of interest \todo{(Fig.~\ref{fig:teaser-overview}D)}. Experts can trace how corresponding neurons in previous layers contribute to the activations of the selected neurons.} 
This is useful for identifying the key 
\newdoc{
learned features} that lead to
\newdoc{
}diverging or merging \jw{of datapaths}. 
In addition, to facilitate the analysis of the diverging/merging patterns for adversarial examples, it is essential to understand how feature maps at each layer contribute to the final prediction. 
In our visualization, the dotted filling style $\vcenter{\hbox{\includegraphics[height=1.5\fontcharht\font`\B]{fig/glyph_dotted.pdf}}}$ is employed to encode 
the contribution from corresponding neurons in previous \jw{layers} 
to the activation of the neurons in the focused area or to the final prediction.
The higher filling 
represents a larger contribution value.

\subsection{Contribution Analysis}
When
\newdoc{
an} expert finds a pattern of interest (e.g., a \todo{critical feature map} in a diverging point or merging point), s/he often \jw{wants} 
to analyze the major cause that 
leads 
to the pattern.
To this end, we \newdoc{have} develop\newdoc{ed} a contribution analysis method to compute the contribution of the previous feature maps to the neuron activation of the feature map of interest (target feature map).

Initially, the contribution analysis is performed based on the whole target feature map.
We formulate this problem as a subset selection problem.
It aims to select a minimum number of feature maps that can maximally preserve the activation of the target feature map.  
This formulation is similar to the datapath extraction discussed in Sec.~\ref{sec:datapath}.
As a result, we also employ the continuous optimization method to select the feature maps and compute the corresponding contribution.
In particular, we replace the first term in Eq.~(\ref{eq:dp-ext-new-3}) with the preservation of the activation of the target feature map:
\begin{equation}
\begin{aligned}
\label{eq:vis-contribution-analysis}
\vecz{opt}^i = & \underset{\vecz{prev}^i\in [0,1]^{n}}{\mathop{\arg \min }} (f(x^i)-f(x^i;\vecz{prev}^i))^2 \\
& + \lambda |\vecz{prev}^i|+ \gamma \sum_{j \in [1, m], j \neq i}||\vecz{prev}^j - \vecz{prev}^i||_2,
\end{aligned}
\end{equation}
where $f(x^i)$ is the 
neuron activation of the target feature map on example $x^i$ and $f(x^i;\vecz{prev}^j)$
\keleiReRe{is the corresponding neuron activation
\newdoc{
in} consideration of}
the previous feature map contributions $\vecz{prev}^j$.
\todo{$m$ is the number of datapaths being analyzed.}
This optimization problem can be solved similarly with our proposed algorithm in Sec.~\ref{sec:datapath}.

\begin{figure}[thbp]
  \centering
  \begin{overpic}[width=0.98\linewidth]{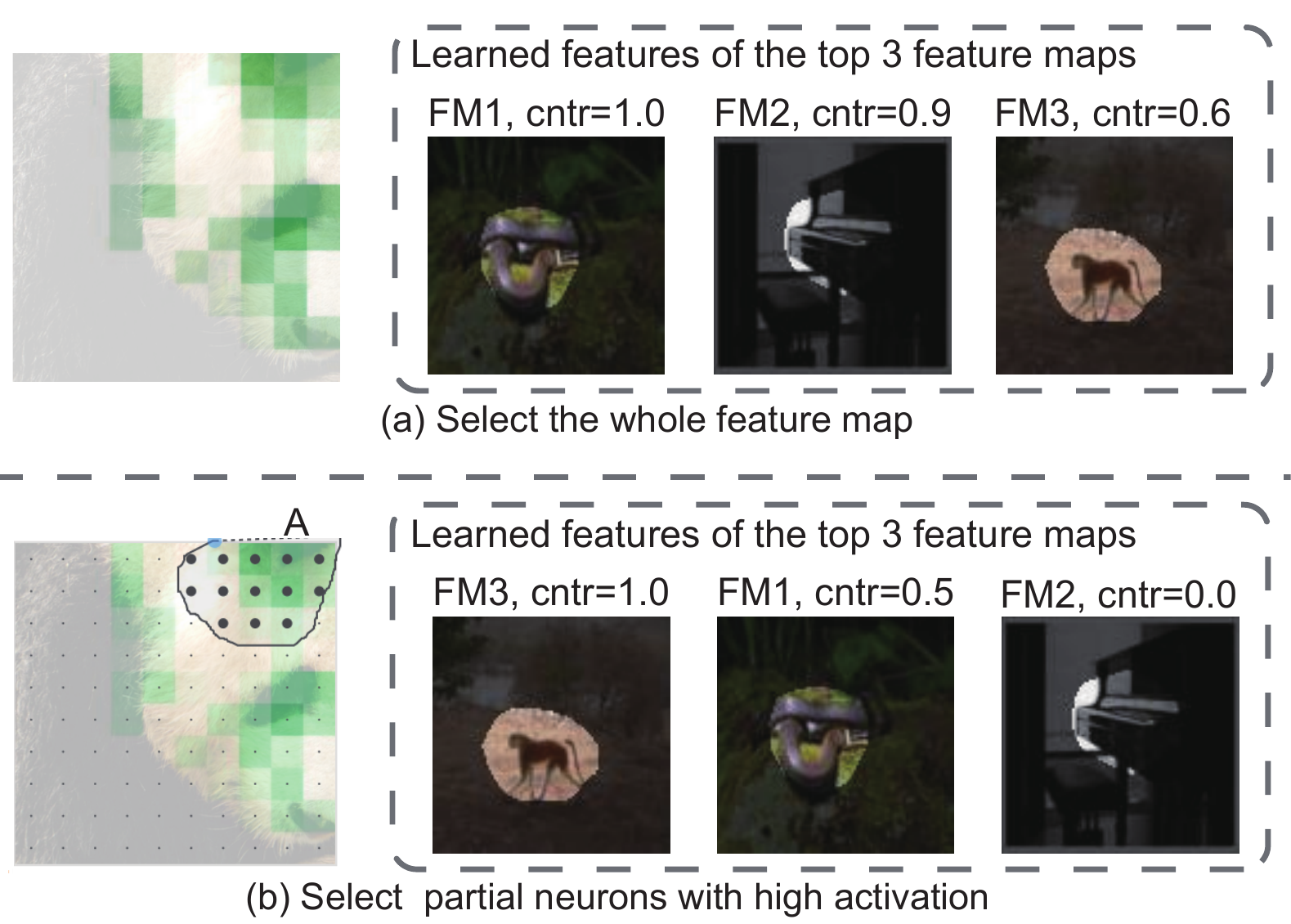}
  \small
  \end{overpic}
  \caption{The top 3 most contributed feature maps identified by considering a) the whole feature map of interest; b) part of the feature map of interest.}
  \label{fig:vis-focus-contribution}
  \vspace{-5mm}
\end{figure}

When using this contribution analysis method to analyze the adversarial noise, we
\newdoc{
discover} an issue. 
As shown in Fig.~\ref{fig:vis-focus-contribution} (a), if all the neurons of the target feature map are considered, some irrelevant feature maps, such as FM1 and FM2, are ranked highest, while the relevant one, FM3, is ranked
\newdoc{
}third. 
This is because, in addition to 
the feature that is misled by the adversarial noise, other irrelevant features with high neuron activation, are also considered. 
These irrelevant features may trigger several irrelevant feature maps and rank them higher.

To tackle this issue, we allow
\newdoc{
}expert\newdoc{s} to only select neurons (Fig.\ref{fig:vis-focus-contribution}A) that are highly activated on the adversarial noise and examine the influence of other feature maps
\newdoc{
on} these selected neurons.
For example, when an expert examines a feature map of interest (e.g., Fig.~\ref{fig:teaser-overview}C),
\keleiRe{
s/he} finds that a certain area of the example is identified as a panda's ear.
S/he then checks the previous layers to investigate the
\newdoc{
reason} why this area
is activated by the neurons.
Therefore, it is useful for the expert to focus on the neurons in this area and check the contribution of the previous feature maps to these selected neurons in the prediction.
The key challenge of this problem is to calculate the contribution of the corresponding neurons in each of the previous feature maps. 
Here the corresponding neurons correspond to the selected neurons in the target feature map.
Similarly, we also aim to select a minimum number of feature maps that can maximally preserve the activation of the selected neurons in the target feature map.  
Accordingly, we change the optimization variable $z_s$ in Eq.~(\ref{eq:vis-contribution-analysis}) from the whole feature map to the corresponding neurons:

\begin{align} 
\label{eq:vis-partial-analysis-fm}
\vecz{opt,p}^i = & \underset{\vecz{prev,p}^i\in [0,1]^{n}}{\mathop{\arg \min }} (f(x^i)-f(x^i;\vecz{prev,p}^i))^2 + \lambda |\vecz{prev,p}^i|\notag\\
& + \gamma \sum_{j \in [1, m], j \neq i}||\vecz{prev,p}^j - \vecz{prev,p}^i||_2,
\end{align} 
where $\vecz{prev,p}=[z^1_{prev,p}, \cdots , z^n_{prev,p}]$.
$z_{prev,p}^k \in [0,1]$ approximates the contribution of the neurons in the focused area of $k$-th feature map to the activation of the selected neurons in the target feature map.

The top 3 most contributed feature maps identified by the new method are shown in (Fig.\ref{fig:vis-focus-contribution} (b), where the most relevant one, FM3, ranks first.


%% file: application.tex
\begin{table*}[bt]
    \centering
    \begin{tabular}{c|c|cccccccccc|c}
    \toprule
    & & \multicolumn{10}{|c|}{Dataset 1} & Dataset 2 \\
    \midrule
       &  & jeep & schooner & banana & pizza & panda & goldfish & rosehip & snake & tusker & sunglass & 100 class (average) \\
    \midrule 
    \textbf{Top-1 Score} & DGR  & 0.000 & 0.038 & 0.039 & 0.059 & 0.051 & 0.006 & 0.000 & 0.025 & 0.000 & 0.029 & 0.011 \\
    & Ours & 0.017 & 0.076 & 0.058 & 0.078 & 0.061 & 0.029 & 0.045 & 0.025 & 0.021 & 0.057 & 0.042\\
    \midrule  
    \textbf{Top-3 Score} & DGR  & 0.033 & 0.114 & 0.107 & 0.176 & 0.111 & 0.041 & 0.061 & 0.101 & 0.042 & 0.095 & 0.064 \\
    & Ours & 0.083 & 0.253 & 0.165 & 0.333 & 0.222 & 0.110 & 0.091 & 0.139 & 0.104 & 0.152 & 0.123 \\
    \midrule  
    \textbf{Top-5 Score} & DGR  & 0.067 & 0.177 & 0.117 & 0.275 & 0.172 & 0.076 & 0.091 & 0.215 & 0.083 & 0.162 & 0.123 \\
    & Ours & 0.133 & 0.468 & 0.204 & 0.647 & 0.404 & 0.174 & 0.136 & 0.291 & 0.167 & 0.248 & 0.209 \\
    \bottomrule
    \end{tabular}
    \caption{The top -1, -3, -5 scores on the 10-class and the 100-class datasets using DGR~\cite{wang2018interpret} and our method for datapath extraction.} 
    \label{tab:QuanExpr_0}
\end{table*}

\section{Evaluation}
\newdoc{
We} first quantitatively evaluated the effectiveness of the proposed constrained datapath extraction method in comparison with a previous state-of-the-art method, the DGR method~\cite{wang2018interpret}.
We then demonstrated through a case study how AEVis helped the analysis of the root cause for misclassification of adversarial examples.
\jing{Expert \caseexpert{1}, one of the two experts \newdoc{who} participated in the evaluation of the previous version of AEVis~\cite{liu2018analyzing}, was invited again to evaluate the usefulness of the new system.}
The CNN used for evaluation is a pretrained
\textbf{ResNet-101}~\cite{He2016_CVPR_residual}, which contains 101 layers and is 
\newdoc{
a} state-of-the-art CNN\newdoc{
} for image classification.

\subsection{Quantitative Analysis}
\label{sec:eval_dp}

\mc{
As there is no ground-truth for datapaths, the effectiveness of the datapath extraction method is measured by the ability to detect the \jw{diverging-merging} patterns between the extracted datapaths.
\jw{Two datasets with different scales were used for this evaluation. One dataset contains all the images of 10 randomly selected classes 
\jw{(shown at the top of Table~\ref{tab:QuanExpr_0})} 
from ImageNet ILSVRC 2012~\cite{russakovsky2015imagenet}. The other contains 100 classes, with 10 randomly selected images in each class.}
We used a state-of-the-art attacking method, namely, \newdoc{the} momentum iterative fast gradient sign method~\cite{NIPS_attack_code,dong2018boosting}, to generate an adversarial example for each image in the datasets.
}
\jw{From the classification results of these adversarial images, for each \jing{class that was mistakenly classified into}, we further sampled $20$ 
target images from the original ImageNet ILSVRC 2012 dataset.}
\jw{Then for each misclassified adversarial image, we constructed $20$ 
triplets of normal source/adversarial/normal target images and extracted datapaths for each triplet.}

As shown in Fig.~\ref{fig:case0_overview}, 
there is a diverging point (\pictag{L}{A}) where the datapath of the misclassified adversarial image gradually deviates from the datapath of the normal source image, and gets closer to that of the normal target image and merges with it at a point (\pictag{L}{E}), resulting in the misclassification.
Such a diverging followed by \newdoc{a} merging pattern (simplified as a \textbf{diverging-merging pattern}) is an important characteristic indicating the misclassification of adversarial images. The ability of the extracted datapaths to reflect such patterns is thus used to evaluate the effectiveness of the datapath extraction method.

\jw{To determine the occurrence of a diverging-merging pattern, we calculated the difference 
\newdoc{
in the} distances between 1) the datapaths of the adversarial and normal source images and 2) the datapaths of the adversarial and normal target images. The difference at layer $i$ is calculated as:}
\begin{equation}
\label{eq:case-pattern-diff}
diff(i) = ||\vecz{opt}^{adv}(i)-\vecz{opt}^{src}(i))||_2 -  ||\vecz{opt}^{adv}(i)-\vecz{opt}^{tar}(i)||_2
\end{equation}
where $\vecz{opt}^{adv}$, $\vecz{opt}^{src}$, and $\vecz{opt}^{tar}$ denote datapaths of the adversarial example, the normal source image that corresponds to the adversarial, and the normal target image, respectively. 

If the distance difference of the last $r$ layers
\newdoc{
continues to increase} towards the end of the model, a diverging-merging pattern is detected. 
Thus, we \newdoc{
count} the number of layers ($n_l$) that 
continuously increasing in the last $r$ layers:
 \begin{equation}
\label{eq:case-pattern-count}
n_l=\sum_{i=m-r+1}^{m} \mathbb{I}(diff(i)>diff(i-1))
\end{equation}
where $\mathbb{I}(\cdot)$ is an indicator function. It equals \newdoc{
}1 if the predicate is true, and 0 otherwise. 
$m$ is the number of layers in the model, \jw{and $r$ is a number recommended by experts to set the minimum length of the diverging-merging pattern \todo{($r=8$ in our experiments)}. 
Then such a pattern is detected when $n_l=r$ and $diff(m)$ reaches the maximum.} 


\jw{
To evaluate the effectiveness of the datapath extraction method, we then defined the following score:} 

\noindent 
\kelei{
\textbf{Top-}\textbf{\emph{K}} \textbf{score}.
For each \jw{misclassified} adversarial example, we sorted the corresponding target images in descending order 
according to their datapath similarity (based on the distance defined in Eq.~(\ref{eq:dp-ext-new-3}), 1/dis) with the adversarial image.
The top-\emph{K} score is calculated as the average number of diverging-merging patterns in the top-K target-images of \jw{each adversarial example.}
}
\jw{The higher the score, the better detection of the diverging-merging pattern in the extracted datapaths, and thus more effective the datapath extraction method.} 

We computed the top-$1$, top-$3$, and top-$5$ scores of our datapath extraction method. For comparison, we also computed the scores for the DGR method~\cite{wang2018interpret}.
\kelei{
Table~\ref{tab:QuanExpr_0} shows 
\jw{the computed scores on the $10$ randomly selected classes in the first dataset. It can be seen} 
that our method performs better than the DGR method on all the classes.
We further computed the scores on the 100 classes in the second dataset, and the average result shown in the last column of Table~\ref{tab:QuanExpr_0} further verified the effectiveness of our method.
}

\subsection{Case Study}
\label{sec:case}

\jw{We invited 
\jing{expert} \caseexpert{1}, to evaluate the usefulness of AEVis.}
\jw{As \caseexpert{1} participated in the aforementioned NIPS 2017 adversarial attack competition, he was interested in using the same \textbf{DEV} dataset from the competition~\cite{NIPS2017_Challenge}. He would like to see whether AEVis could help him gain a better understanding of the misclassification of adversarial examples.} 
\jw{The \textbf{DEV} dataset contains 1000 images of different classes, } 
\jw{and for each image, we generated an adversarial image using the non-targeted attacking method developed by the winning team~\cite{NIPS_attack_code,dong2018boosting}. }

\begin{figure}[htbp]

  \centering
  \begin{overpic}[width=0.98\linewidth]{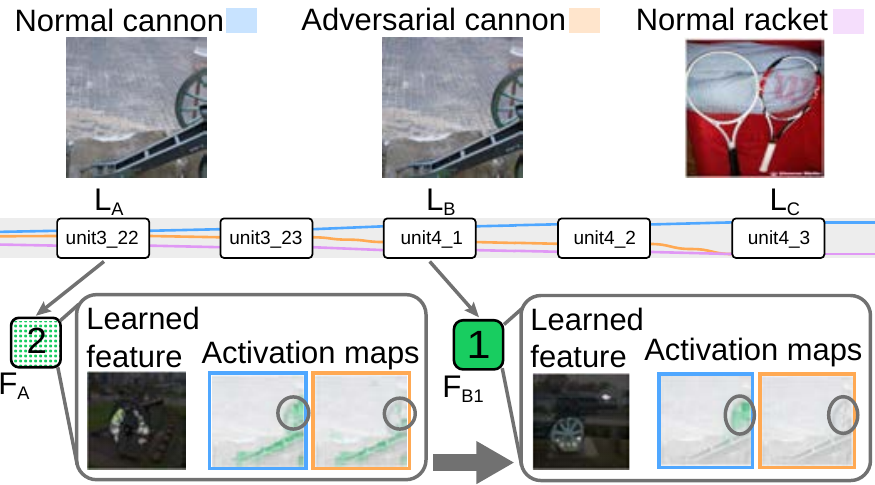}
  \small
  \end{overpic}
  \caption{
  \kelei{Diverging analysis for an adversarial cannon image.}
  }
  \label{fig:case2-diverging}
\end{figure}

\jw{To facilitate the analysis,
\keleiReRe{we calculated an adversarial score}
for each adversarial image using the method in~\cite{pang2018towards}. A higher score means \newdoc{a} more
obvious \newdoc{
}adversarial example. These scores, together with the classification results, 
were presented to \caseexpert{1} for him to select suitable adversarial examples for analysis. \caseexpert{1} was interested in misclassified adversarial examples with medium scores, as 
he commented, `Less obvious examples often contain subtle changes with big influence'.} 
After examining these uncertain adversarial examples, 
\jing{\caseexpert{1} selected two images for further investigation:
an image of a panda head 
\newdoc{
that had been} misclassified as a guenon monkey (\pictag{I}{2} in Fig.~\ref{fig:teaser-overview}), and an image of a cannon misclassified as a racket (Fig.~\ref{fig:case2-diverging}).}


\jing{\subsubsection{Panda image}}
\label{sec:panda}
\kelei{
To \jw{find and} understand the root cause of this misclassification, \caseexpert{1} 
\jw{selected the adversarial panda image (\pictag{I}{2} in Fig.~\ref{fig:teaser-overview}), the normal panda image (\pictag{I}{1} in Fig.~\ref{fig:teaser-overview}), and $10$ normal guenon monkey images (\pictag{I}{3} in Fig.~\ref{fig:teaser-overview}) in the AEVis system.} 
The datapaths of normal and adversarial panda images and a representative monkey image were then automatically extracted for further analysis.
\mcrev{In particular, the representative monkey image was selected from $10$ randomly sampled monkey images, \newdoc{among which it had} the highest datapath similarity with the adversarial panda image (Eq.~\ref{eq:dp-ext-new-3}).
}
}

  

\begin{figure}[bp]
  \centering
  \begin{overpic}[width=0.99\linewidth]{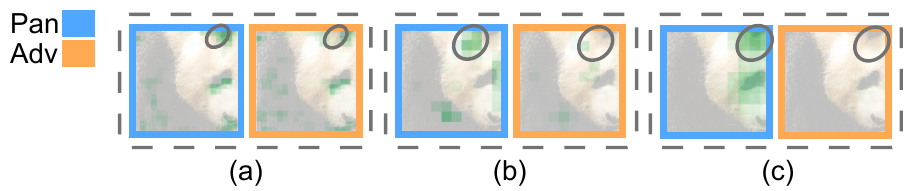}
  \small
  \end{overpic}
  \caption{
  \kelei{Activation of feature maps in (a) \pictag{L}{A}, (b) \pictag{L}{D} and (c) \pictag{L}{E}.}
  }
  \label{fig:f_d_e}
\end{figure}

\begin{figure*}[tb]
  \centering
  \vspace{2mm}
  \scriptsize
  \begin{overpic}[width=0.98\linewidth]{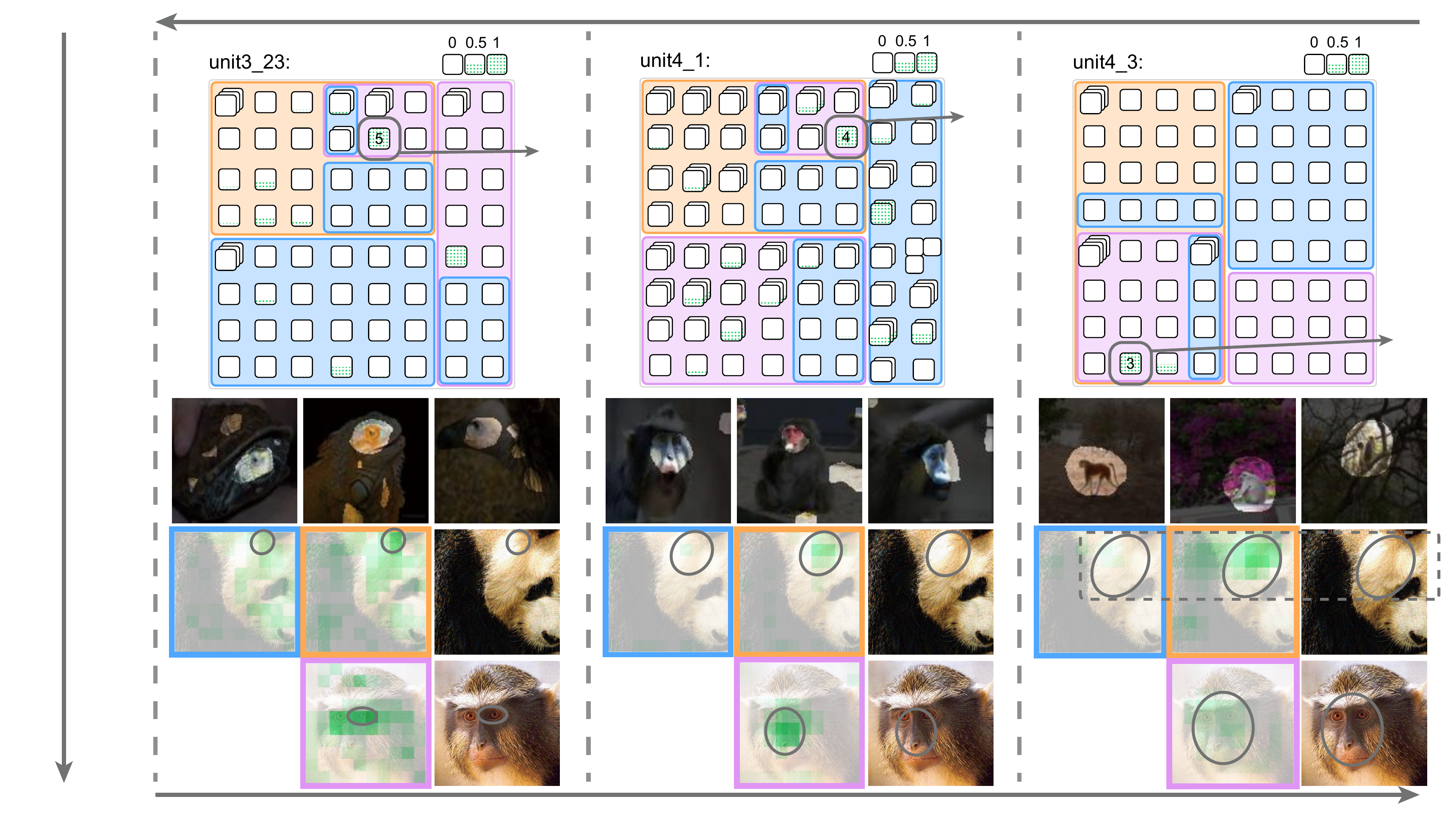}
  \put(42.5, 55.5){\textsf{Direction for merging analysis}}
  \put(2.5, 18){\begin{sideways}\textsf{Analysis for a single feature map}\end{sideways}}
  \put(7, 37){\begin{sideways}\textsf{(i) Layer-level}\end{sideways}}
  \put(9, 37){\begin{sideways}\hspace{2mm}\textsf{visualization}\end{sideways}}
  \put(7, 20.5){\begin{sideways}\textsf{(ii) Learned}\end{sideways}}
  \put(9, 20.5){\begin{sideways}\hspace{3mm}\textsf{features}\end{sideways}}
  \put(7, 7){\begin{sideways}\textsf{(iii) Activation}\end{sideways}}
  \put(9, 7){\begin{sideways}\hspace{5mm}\textsf{maps}\end{sideways}}
   \put(95.8, 32.5){$\textsf{F}_\textsf{E}$}
   \put(66.2, 48){$\textsf{F}_\textsf{C2}$}
   \put(37, 45.5){$\textsf{F}_\textsf{B}$}
    \put(72, 30.5){$\textsf{L}_\textsf{E}$}
   \put(42.2, 30.5){$\textsf{L}_\textsf{C}$}
   \put(12.5, 30.5){$\textsf{L}_\textsf{B}$}
   \put(99.3, 18){\textsf{A}}
  \put(11.5, 3){\textsf{(a)}}
  \put(41.2, 3){\textsf{(b)}}
  \put(71, 3){\textsf{(c)}}
  \put(42, -0.5){\textsf{Direction of dataflow in the model}}
  \end{overpic}
  \vspace{2mm}
\caption{
\kelei{\caseexpert{1}'s analysis process from the deep layer \pictag{L}{E} to the shallow layer \pictag{L}{B} 
\newdoc{
in order to find} the major cause of merging.}}
  \label{fig:case-merging-overview}
\end{figure*}

\noindent \textbf{\normalsize Overview.}
\kelei{
\jw{The system first displayed the datapaths at the network-level (Fig.~\ref{fig:teaser-overview} (b)).
The distances among the three datapaths disclose the diverging-merging patterns through the layers.}
%
\jw{Following the dataflow in the overview, \caseexpert{1}} found that the datapath of the adversarial panda image 
\jw{began to deviate 
from the datapath of the normal panda image at layer \caselayer{A} (Fig.~\ref{fig:case0_overview}), gradually got closer to the datapath of the monkey image, and finally merged into it at \caselayer{E} (Fig.~\ref{fig:case0_overview}).} 
\jw{From \caselayer{A} to \caselayer{E} are} the layers where the predictive behaviors of neurons were misled by the adversarial noise. 
To \jw{better} understand the 
\jw{working mechanism} 
of adversarial noise, 
he then analyzed the misclassification from two aspects: 
the \textbf{diverging process} of the adversarial image from the normal source image (panda), and 
the \textbf{merging process} of the adversarial image into the normal target images (monkey).
}

\noindent \textbf{\normalsize Diverging analysis}.
To analyze which feature map\mc{s} \jw{in the datapath of the adversarial example} \mc{were} critical for the divergence,  \caseexpert{1} first expanded \jw{layer \caselayer{C} where the divergence became noticeably large.}
\kelei{
The encoded value of each feature map 
was set as the activation difference between \jw{the normal \mc{source}} 
and \jw{adversarial images.}
\jw{A large difference indicates that the feature map
\newdoc{
has detected} its learned features in the normal source image but not in the adversarial example.}
\jw{With this understanding, \caseexpert{1}} directly checked the feature map \jw{\pictag{F}{C1}}, which had the largest activation difference \jw{in \pictag{L}{C}}. 
\doc{
By examining the learned features (Fig.~\ref{fig:teaser-overview}A) of this feature map, he discovered that the neurons in this feature map were trained to detect a black circle pattern that resembled an ear or an eye of a panda (Fig.~\ref{fig:teaser-overview}B and C).
\jw{Such a pattern is \mc{one of} the unique characteristics of a panda and is thus critical for its classification.}
\jw{Then looking at the activation maps (Fig.~\ref{fig:teaser-overview}C), \caseexpert{1} noticed that the neurons covering the ear area were correctly activated for the normal panda image, indicating a successful detection of this critical pattern. However, the same neurons were not activated for the adversarial example. \caseexpert{1} considered this was an important reason for the misclassification.}
}
}

\jw{To analyze the root cause for this failed detection,} 
\kelei{
\caseexpert{1} selected \newdoc{
the} neurons covering the ear area to form an \todo{area of interest} (Fig.~\ref{fig:teaser-overview}D) for
\jw{a closer examination.} 
He suspected the failed detection was influenced by the feature maps from previous layers.
He then set the value encoded in each feature map as the `contribution' to the selected one, i.e. the \todo{area of interest} in \casefm{C1}, 
\jw{and expanded \pictag{L}{A}, the layer at the beginning of the diverging process.} 
\jw{In the treemap-based visualization at \pictag{L}{A}, \caseexpert{1} found that feature map \casefm{A} (Fig.~\ref{fig:teaser-overview} (b)) which had the largest contribution to the activation difference in \casefm{C1}.} 
By examining its learned features (Fig.~\ref{fig:teaser-overview}E), \caseexpert{1} confirmed that 
\jw{it was trained for low-level detection of black-white boundaries.}
\jw{The activation maps (Fig.~\ref{fig:f_d_e} (a)) showed that this feature was detected in the normal panda image but not in the adversarial example. \caseexpert{1} speculated that this failed detection led to the failed detection of the panda's ear in \casefm{C1}, and finally led to the failed classification of the adversarial example as a panda.}
}

\jw{To confirm his speculation, \caseexpert{1} repeated the analysis on \pictag{L}{D} and \pictag{L}{E}. 
\newdoc{
In these two layers, he} selected \newdoc{
}several feature maps with a big activation difference between the normal and adversarial images.
From the activation maps, he found \newdoc{
}more significant
\newdoc{
misses in the} detection of critical patterns in the adversarial example (Fig.~\ref{fig:f_d_e} (b), and (c)). Selecting the area of interest and tracing the contribution back to \pictag{L}{A}, \caseexpert{1} identified the same feature map \pictag{F}{A} as the biggest contributor to the failed detection in \pictag{L}{D} and \pictag{L}{E}.
At this point, \caseexpert{1} was convinced that the missing detection of the \newdoc{
black-white} boundary in \pictag{F}{A}
\newdoc{
was} the root cause for the failed detection of critical patterns in higher levels
\newdoc{
that} finally led to the failed classification of the adversarial example as a panda.}

\noindent \textbf{\normalsize Merging analysis}.
\jw{After analyzing the 
\newdoc{
reason why} the adversarial example 
\newdoc{
was not} classified as a panda, \caseexpert{1} turned his attention to why it was classified as a monkey.}
\jw{He suspected that the same region that led to the failed \newdoc{classification of panda} actually contributed to 
\newdoc{
its} classification as a monkey.}
\jw{Therefore, he \newdoc{
retained the same}} \todo{area of interest} 
and expanded 
\newdoc{
}layer \caselayer{E} (Fig.~\ref{fig:case0_overview}), the merging point for the datapaths of the adversarial and the monkey images.
To find the feature maps that 
\newdoc{
had} the main contribution to the misclassification, 
he set the encoded value of each feature map as the `contribution' to the prediction of the adversarial panda image 
\jw{and identified \newdoc{the} feature map (\casefm{E} in Fig.~\ref{fig:case-merging-overview} (c)) with the largest contribution.}
After examining its learned features (Fig.~\ref{fig:case-merging-overview} (c)(ii)),
\caseexpert{1} discovered that it was trained to detect monkeys in various situations.
\jw{Comparing the activation maps of the adversarial example and normal monkey image (Fig.~\ref{fig:case-merging-overview} (c)(iii)), he found that the monkey face was activated in the monkey image as expected.
However, it 
\newdoc{
was} hard to explain the activation at the top part of the adversarial example. Intuitively, there were no indications of \newdoc{a} monkey in that part.}
Thus, \caseexpert{1} decided to trace back to the lower levels to seek more clues.

\jw{Guided by the larger activation on the activation map of the adversarial example, \caseexpert{1} first adjusted the \todo{area of interest} to include the most activated neurons at this layer (Fig.~\ref{fig:teaser-overview}D), and then analyzed the contributions to \pictag{F}{E} from the feature maps in previous layers.} 
\jw{In layer \caselayer{C}, he found feature map \pictag{F}{C2} (Fig.~\ref{fig:case-merging-overview} (b)), which 
\newdoc{
had} the highest contribution to \pictag{F}{E}.} 
\jw{\newdoc{
After} a closer look, the neurons in \pictag{F}{C2} seemed to detect the face of a monkey}  (Fig.~\ref{fig:case-merging-overview} (b)(ii)). 
\jw{Activation on the monkey image was correctly located in the middle of the monkey face, but in the adversarial panda image, \newdoc{it} was again located on the top right part as was \newdoc{the case} in \pictag{F}{E} (Fig.~\ref{fig:case-merging-overview}\keleiRe{
(c)}). }
\jw{`Are there patterns of a monkey face?' With this question \newdoc{in mind}, \caseexpert{1} carefully compared the adversarial example and the monkey image (Fig.~\ref{fig:case-merging-overview} (b)(iii)). He found that for the adversarial example, in the activated region, the dark strip with two lighter patches 
\newdoc{
on either side} did resemble \newdoc{the look of} a monkey's nose with lighter cheeks 
\newdoc{
next to it}.}

\begin{figure}[thbp]
  \centering
  \begin{overpic}[width=0.8\linewidth]{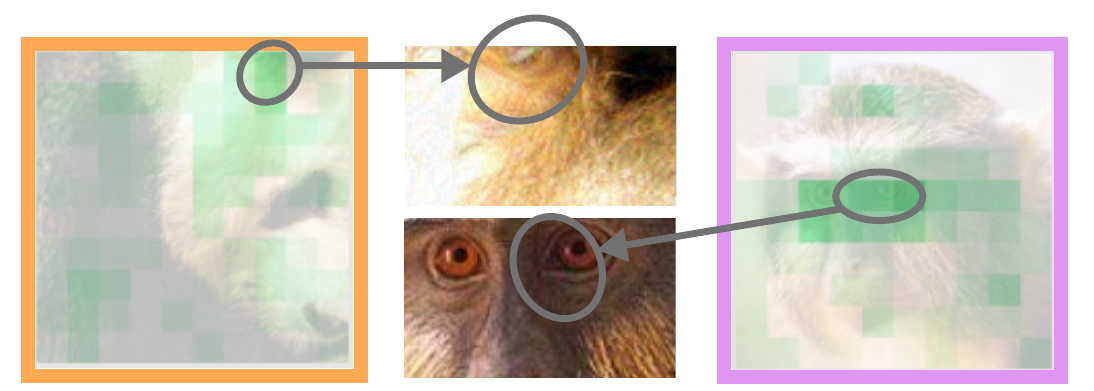}
  \small
  \end{overpic}
  \caption{A small region with the appearance of an eye was detected at the top right corner of the adversarial panda image.}
  \label{fig:case-end}
\end{figure}

\jw{However, the same pattern was \newdoc{
present} in the normal panda image, but it was not detected by the neurons in \pictag{F}{C2}.} 
\jw{\caseexpert{1} thus wanted to investigate \keleiReRe{the} more fundamental cause 
\newdoc{
for} the detection.} 
\jw{Again, he adjusted the \todo{area of interest} according to the activation map and expanded} 
the previous layer \caselayer{B} (Fig.~\ref{fig:case0_overview}). 
\jw{He identified \newdoc{that} feature map \pictag{F}{B} (Fig.~\ref{fig:case-merging-overview} (a)) \newdoc{
}was in the datapath intersection for both the adversarial and monkey images and had the largest contribution to \casefm{C2}.} 
Examining the learned features of this feature map (Fig.~\ref{fig:case-merging-overview} (a)(ii)),
\caseexpert{1} found that the neurons inside were trained to identify the eyes of different animals. 
Further inspecting the activation of the three images, he found that \jw{a small region with the appearance of an eye was detected} at the top right corner of the adversarial panda image. 
(Fig.~\ref{fig:case-end}).
\jw{And compared with the normal image, this seemed attributed to the added adversarial noise.} 
\jw{In addition, the position of the `eye' combined with the position of the `nose' detected in \pictag{F}{E} resembled the layout of a real monkey face.} 
\jw{\newdoc{
At} this point, \caseexpert{1}} 
figured out the effect of the adversarial noise.
The top right corner of the adversarial panda image had some similarities with a monkey's face.
In particular, the imperceptible adversarial noise misled the model to detect a monkey's eye first, 
\jw{then the subtle changes in image layout misled the model to detect a monkey's face. It was like a domino effect and finally led to the misclassification of the adversarial example as a monkey.}

\noindent \textbf{\normalsize Summary.} From the above analysis, \caseexpert{1} summarized two effects of the adversarial noise. 
The first one was that 
\jw{the outline of the panda's ear was affected by the noise, which led to the failed detection of the ear and} 
resulted in the large decrease of the predicted probability of the panda class.
The second one is that the adversarial noise misled the model to detect a monkey's eye in the same region, which further led to the detection of a monkey's face.
Then the probability of monkey class largely increased and resulted in the final misclassification.

\jing{\subsubsection{Cannon image}}
\label{sec:cannon}
\jing{\caseexpert{1} carried out a similar examination 
\newdoc{on} the adversarial cannon image. Examining the datapaths of the normal and adversarial cannon images and a representative racket image (Fig.~\ref{fig:case2-diverging}), \caseexpert{1} first identified the diverging and merging points (\caselayer{B} and \caselayer{C} in Fig.~\ref{fig:case2-diverging}).
To analyze the root cause for divergence, \caseexpert{1} followed the same process as above. He first discovered the activation difference between the normal and adversarial images on \newdoc{
}feature map \casefm{B1} that was trained to detect the wheel of a cannon (Fig.~\ref{fig:case2-diverging}). He then traced the failed detection to \newdoc{
}feature map \casefm{A} in \caselayer{A}, where the wheel shafts were not detected in the adversarial example (Fig.~\ref{fig:case2-diverging}). \caseexpert{1} speculated the added noise blurred the edges of the shafts which resulted in the failed detection, and further led to the failed detection of the wheel, and finally the misclassification.}

\begin{figure}[htbp]
  \centering
  \begin{overpic}[width=0.95\linewidth]{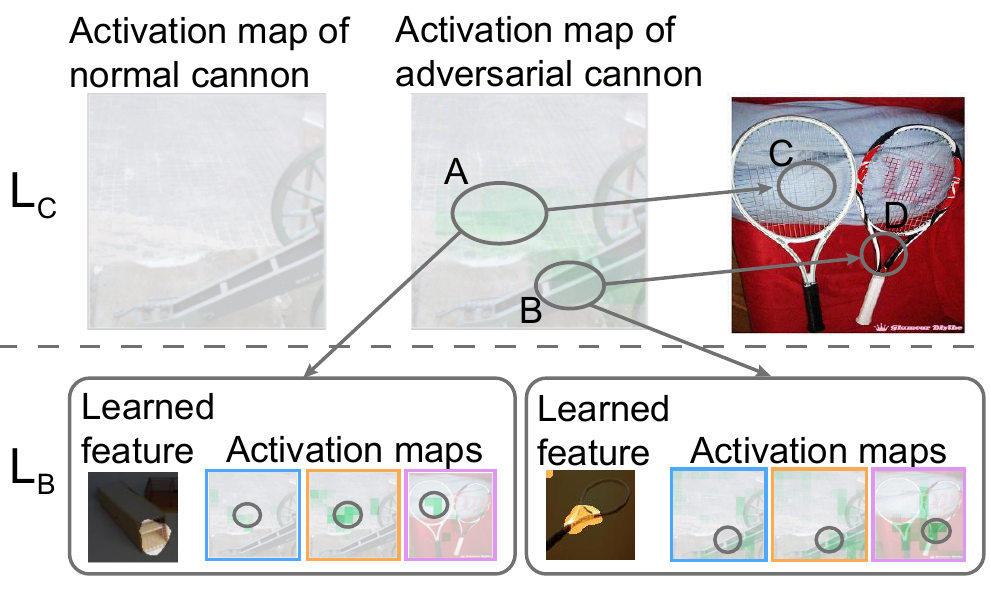}
  \small
  \end{overpic}
  \caption{
  \kelei{Merging analysis for adversarial cannon image.}
  }
  \label{fig:case2-merging}
\end{figure}


\jing{To understand why the adversarial example was misclassified as a racket, \caseexpert{1} turned his attention to the merging point \caselayer{C}, and identified the feature map \casefm{C} which had the largest contribution to the misclassification (Fig.~\ref{fig:case2-merging}). Comparing the activation maps (Fig.~\ref{fig:case2-merging}\caselayer{C}), he found two regions (Fig.~\ref{fig:case2-merging}A and B) that were wrongly activated in the adversarial cannon image. Selecting each region and tracing back to layer \caselayer{B}, \caseexpert{1} noticed the feature maps that contributed the most to each of the regions were trained to detect `net' and `racket throat' respectively (Fig.~\ref{fig:case2-merging}C and D).
There are similarities between the streaks on the ground and a racket \mcrev{net}, 
and between the gun mount and a racket throat. However, the added noise in the adversarial image 
\newdoc{
creates a} stronger activation in the two feature maps (Fig.~\ref{fig:case2-merging}\caselayer{B}). \caseexpert{1} thus speculated that the stronger activation together with the failed detection of the wheel misled the model to detect the \mcrev{net} and the throat of a racket in adjacent regions, which finally led to the misclassification of the adversarial image. 
}

%% file: discussion.tex
\section{Discussion}
\label{sec:discussion}

AEVis can \jw{effectively} 
illustrate the prediction mechanism of adversarial examples and help discover the \jw{root cause} 
\doc{that leads} to incorrect predictions. 
However, it \jw{still} 
has several limitations, which may shed light on future research \doc{directions}.


\noindent \textbf{Time complexity.} The datapath extraction usually takes a few minutes and is computed offline. \mcrev{The contribution analysis is the only part that cannot be pre-computed because the contribution is calculated based on the selected neurons.
It usually takes about $5$ seconds to calculate 
\newdoc{
the} contribution using SGD to solve Eq.~(\ref{eq:vis-partial-analysis-fm}).
\newdoc{
To accelerate the process}, we can use the quadratic approximation
\newdoc{
from} our previous work~\cite{liu2018analyzing}, which
\newdoc{
is faster} (computation time $<1s$) but
\newdoc{
less accurate.}
\newdoc{
Since} our target users (machine learning experts) focus more on analysis accuracy, the SGD-based solution is set as default.
\newdoc{
Users} can switch to the approximated contribution analysis in the interface.
}

\noindent \textbf{\normalsize Visual scalability}.
We have demonstrated that AEVis is able to analyze a state-of-the-art CNN (ResNet101), which has 101 layers and is much deeper than traditional CNNs (e.g., VGG-Net).
More recently, \jw{deeper CNNs with thousands of layers~\cite{He2016_CVPR_residual} have been developed.} 
When handling such deep \jw{neural networks, the layers of interest at low levels of the hierarchy \jw{are difficult to} 
fit in one screen, even with the help of our segmented DAG.} 
\keleiReRe{A possible solution to alleviate this issue is to}
employ a mini-map to help the expert track \doc{the} current viewpoint, which has \newdoc{
}proven effective in TensorFlow~\cite{wongsuphasawat2018visualizing}.



\mc{
Currently, we utilize a river-based visual metaphor to illustrate the diverging and merging patterns.
The layout of the datapaths is calculated 
\newdoc{
using} a rule-based method (Sec.~\ref{sec:vis-net}).
Such a design 
\keleiReRe{echoes}
the most common analytical task 
\newdoc{
when} three datapaths 
\newdoc{
need} to be compared (adversarial examples, the normal source examples, and the normal target examples).
If more datapaths are to be analyzed, an optimization-based layout method can be applied.
For example, we can minimize the mean-square-error between the vector of real datapath distances and their screen distances with a constraint \newdoc{so} that the order of real datapath distances
\keleiReRe{is} maintained. 
The above optimization problem is convex (convex functions over convex sets) and guaranteed to achieve a global minimum.
As we have not observed such needs, we leave this method in the discussion here. 
Apart from the river-based visualization, the treemap-based visualization in the layer level is the other factor that limits the 
\newdoc{
ability to analyze} a lot of datapaths.
The intuitive treemap-based design is suitable for comparing several datapaths~\cite{alsallakh2014visualizing} and has been 
\newdoc{
proven} effective in the case studies.
We can further improve its scalability by adopting \newdoc{a} less intuitive but more scalable set \newdoc{of} visualization techniques, such as PowerSet~\cite{alsallakh2017powerset}.}


\noindent \textbf{\normalsize Generalization}.
AEVis aims 
\newdoc{
to analyze} the adversarial examples for CNNs because most research on adversarial attacks focuses on generating adversarial images for CNNs.

In addition to attacking CNNs, there are several initial attempts to attack other types of DNNs~\cite{Akhtar2018_threat}, such as recurrent neural networks (RNNs), autoencoders (AEs), and deep generative models (DGMs).
In these types of DNNs, \jw{there are also neurons that are critical for predictions.}
For example, Ming et al.~\cite{ming2017understanding} demonstrated that some neurons in an RNN were critical for predicting the sentiment of a sentence, such as the neurons for detecting positive/negative words.
Such neurons and their connections form a datapath for an RNN.
Thus, 
AEVis can be extended to help understand the root cause of adversarial examples for these DNNs.
\jw{The main extension \newdoc{required} is the development of} suitable datapath extraction and visualization methods for different types of DNNs.
For example, to visualize the datapath of RNNs, we can first unfold the architecture of an RNN to a DAG~\cite{Goodfellow2016}, and then employ a DAG layout algorithm to calculate the position of each unfolded layer.

In addition to images, \jw{there are adversarial attacks on other types of data~\cite{Akhtar2018_threat}}, 
such as adversarial documents and adversarial videos.
To generalize AEVis to \jw{different} 
types of data, we need to change the visual hint for neurons (learned features and activation maps) according to the target data type.
For example, \doc{when} analyzing adversarial documents, we can use a word cloud to represent the `learned feature' of a neuron~\cite{ming2017understanding}, 
\jw{and} select the keywords that strongly activate the neuron.

%% file: conclusion.tex
\section{Conclusion}
We have presented a robustness-motivated visual analysis
\jw{tool, AEVis, to help} 
machine learning experts \jw{investigate the prediction process and understand the root cause 
\newdoc{
of} incorrect predictions of adversarial examples.} 
\jw{The visualization at multiple levels, together with the constrained datapath extraction, allows efficient identification of critical layers from datapaths' diverging-merging patterns and 
\newdoc{
}critical neurons from the activation maps. The contribution analysis and the rich interactions further enable users to trace the root cause 
\newdoc{
of the} misclassification of adversarial examples.}
 \jw{We conducted a quantitative experiment to evaluate the datapath extraction method and a representative case study with an expert to demonstrate the usefulness of AEVis 
 \newdoc{
 in} explaining the reasons behind \newdoc{the} misclassification of adversarial examples.}

\jw{There are several directions 
\newdoc{
we could} follow in our future 
\newdoc{
research}. First, based on the discovered root cause for misclassification, an interesting and important step forward is to develop targeted defense solutions. We will continue working with \newdoc{
}machine learning experts 
\newdoc{
to} explore an effective route from discovered cause to targeted solutions for developing more adversarial robust DNN models. 
Second, complementary to developing defense solutions, another avenue is to detect potential adversarial examples online and remove them from further processing. A set of streaming visualizations that can incrementally integrate the incoming log data with existing data is the key to online monitoring.
Third, as discussed in Sec.~\ref{sec:discussion}, an interesting direction is to generalize AEVis to analyze the noise robustness of other types of DNNs, and to \keleiReRe{tackle} other types of data. 
\newdoc{
Improving} the scalability to deeper DNNs and the visualization of more datapaths is also 
\newdoc{
an area of future interest.}
Different datapath extraction algorithms and suitable visualization designs would be 
\newdoc{interesting} research topics.
\newdoc{
}}
